\documentclass[sigconf]{acmart}
\AtBeginDocument{%
  \providecommand\BibTeX{{%
    \normalfont B\kern-0.5em{\scshape i\kern-0.25em b}\kern-0.8em\TeX}}}

\copyrightyear{2023}
\acmYear{2023}
\setcopyright{acmlicensed}\acmConference[SIGIR '23]{Proceedings of the 46th International ACM SIGIR Conference on Research and Development in Information Retrieval}
\acmBooktitle{Proceedings of the 46th International ACM SIGIR Conference on Research and Development in Information Retrieval (SIGIR '23)}
\acmPrice{15.00}
\acmDOI{10.1145/3539618.3591679}
\acmISBN{978-1-4503-9408-6/23/07}

\usepackage{multirow}
\usepackage{graphics}
\usepackage{subcaption}
\usepackage{bbding}
\usepackage{enumitem}
\usepackage{balance}
\usepackage{hyperref}

\begin{document}

\title{Ensemble Modeling with Contrastive Knowledge Distillation for Sequential Recommendation}

\author{Hanwen Du}
\email{hwdu@stu.suda.edu.cn}
\affiliation{%
  \institution{Soochow University}
  \city{Suzhou}
  \state{Jiangsu}
  \country{China}
}

\author{Huanhuan Yuan}
\email{hhyuan@stu.suda.edu.cn}
\affiliation{%
\institution{Soochow University}
\city{Suzhou}
\state{Jiangsu}
\country{China}
}

\author{Pengpeng Zhao$^{^*}$}
\thanks{$^*$Corresponding author.}
\email{ppzhao@suda.edu.cn}
\affiliation{%
\institution{Soochow University}
\city{Suzhou}
\state{Jiangsu}
\country{China}
}

\author{Fuzhen Zhuang}
\email{zhuangfuzhen@buaa.edu.cn}
\affiliation{%
\institution{Beihang University}
\city{Beijing}
\country{China}
}

\author{Guanfeng Liu}
\email{guanfeng.liu@mq.edu.au}
\affiliation{%
\institution{Macquarie University}
\city{Sydney}
\country{Australia}
}

\author{Lei Zhao}
\email{zhaol@suda.edu.cn}
\affiliation{%
\institution{Soochow University}
  \city{Suzhou}
  \state{Jiangsu}
  \country{China}
}

\author{Victor S. Sheng}
\email{victor.sheng@ttu.edu}
\affiliation{%
\institution{Texas Tech University}
  \city{Lubbock}
  \state{Texas}
  \country{USA}
}

\fancyhead{}

\begin{abstract}
Sequential recommendation aims to capture users' dynamic interest and predicts the next item of users' preference. Most sequential recommendation methods use a deep neural network as sequence encoder to generate user and item representations. Existing works mainly center upon designing a stronger sequence encoder. However, few attempts have been made with training an ensemble of networks as sequence encoders, which is more powerful than a single network because an ensemble of parallel networks can yield diverse prediction results and hence better accuracy. In this paper, we present \textbf{E}nsemble \textbf{M}odeling with contrastive \textbf{K}nowledge \textbf{D}istillation for sequential recommendation (\textbf{EMKD}). Our framework adopts multiple parallel networks as an ensemble of sequence encoders and recommends items based on the output distributions of all these networks. To facilitate knowledge transfer between parallel networks, we propose a novel contrastive knowledge distillation approach, which performs knowledge transfer from the representation level via Intra-network Contrastive Learning (ICL) and Cross-network Contrastive Learning (CCL), as well as Knowledge Distillation (KD) from the logits level via minimizing the Kullback-Leibler divergence between the output distributions of the teacher network and the student network. To leverage contextual information, we train the primary masked item prediction task alongside the auxiliary attribute prediction task as a multi-task learning scheme. 
Extensive experiments on public benchmark datasets show that EMKD achieves a significant improvement compared with the state-of-the-art methods. Besides, we demonstrate that our ensemble method is a generalized approach that can also improve the performance of other sequential recommenders. Our code is available at this link: \url{https://github.com/hw-du/EMKD}.
\end{abstract}




\maketitle

\section{Introduction}

Recommender systems predict users' interest based on historical interactions. Due to the evolving nature of a user's interest, the task of sequential recommendation views historical interactions as a sequence and aims to model the users' dynamic interest. A typical sequential recommendation model captures such dynamic interest via a sequence encoder, which is a deep neural network that generates the hidden representations of users and items.
\begin{figure}
\centering
\includegraphics[ width=0.8\linewidth]{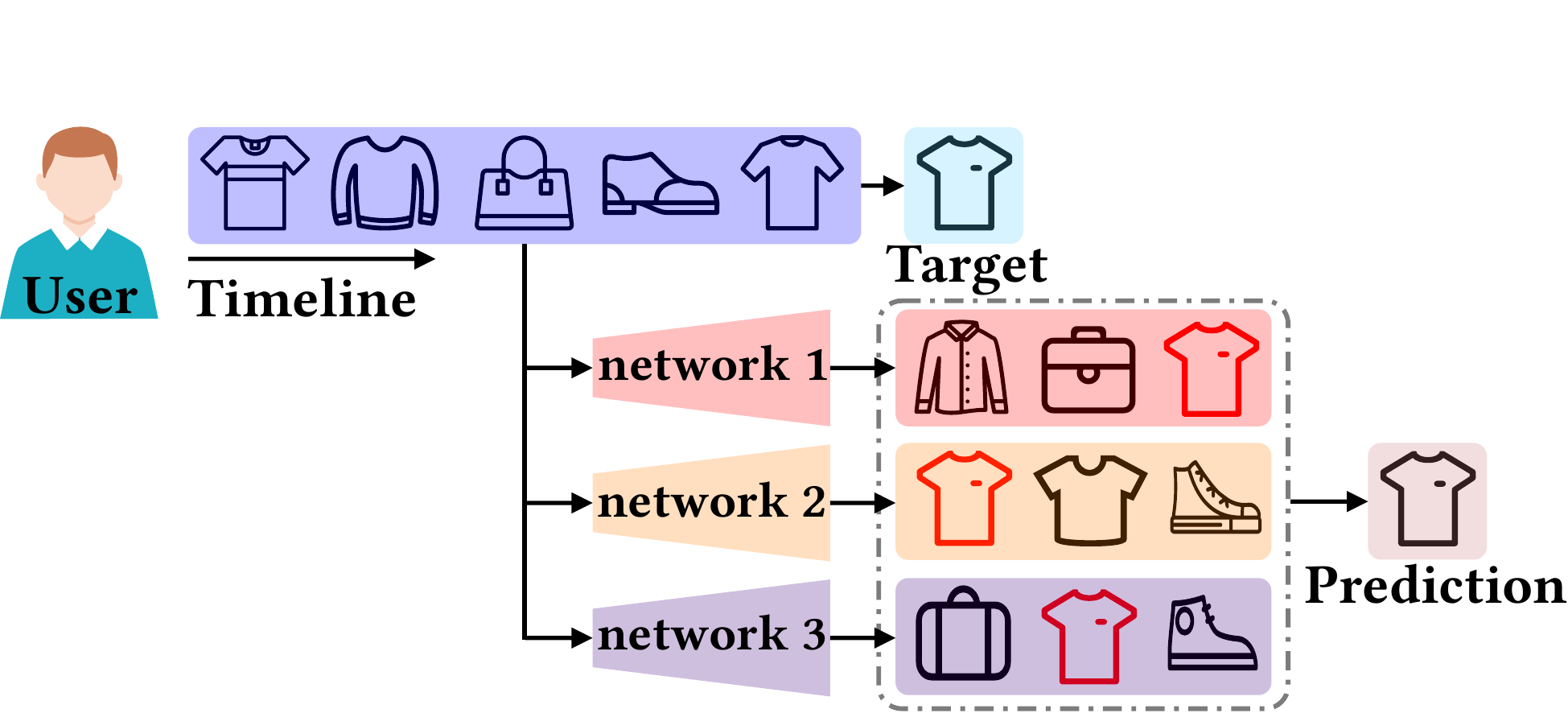}
\caption{An illustration of ensemble modeling for sequential recommendation. Three parallel networks make different predictions based on users' historical interactions. Although each individual network is unable to make an accurate prediction, combining the predictions of these networks together will get the correct result.}\label{ensemble}
\end{figure}Recommendations are made based on the information encoded in the hidden representations. Various types of deep neural networks have been shown as effective sequence encoders in capturing the users' dynamic interest, such as Recurrent Neural Network (RNN) \cite{srnn2016}, Convolutional Neural Network (CNN) \cite{tang2018caser}, unidirectional Transformer \cite{kang18attentive}, and bidirectional Transformer \cite{Sun2019bert}.

To accurately capture the users' dynamic interest, most existing works try to improve the architecture of a single sequence encoder in the hope that a diversity of behavioral patterns, such as skip behavior \cite{tang2018caser} and disordered sequence \cite{Sun2019bert}, can be successfully detected by the sequence encoder. However, an alternative approach---training multiple networks as an ensemble of encoders---is less explored in sequential recommendation. A single deep neural network can only converge to a local minimum, and the number of possible local minima grows exponentially with the number of parameters \cite{localminima}. Therefore, it is hardly possible that two networks with different initializations will converge to the same local minimum even if they share the same architecture \cite{snapshotensemble}.
Such a property can bring benefits, because networks converging to different local minima will make different predictions. Although each prediction can only achieve a certain level of accuracy rate, a diversity of predictions combined together can significantly improve the overall accuracy rate. 
For example, in Figure \ref{ensemble}, parallel networks with different initializations will generate a different list of candidate items as the prediction results. It might be hard to get the correct result if we only consider the prediction from one single network. But if we consider predictions from all these networks and choose the most popular item, it will be much easier to get the correct result. 
\begin{table}
    \centering
    \footnotesize
    \caption{Performance comparison (NDCG@10) between the original model and the ensemble models. We independently train two parallel networks initialized with different random seeds and compare the performance with the original model.}
    \label{simpleensemble}
  \setlength{\tabcolsep}{2pt}{
    \begin{tabular}{lccccccc}
    \toprule
         \multirow{2}[0]{*}{Model}&\multicolumn{2}{c}{GRU4Rec}&\multicolumn{2}{c}{Caser}&\multicolumn{2}{c}{SASRec}\\\cmidrule(lr){2-3}\cmidrule(lr){4-5}\cmidrule(lr){6-7}
         &Original&Ensemble(2$\times$)&Original&Ensemble(2$\times$)&Original&Ensemble(2$\times$)\\\midrule
         Beauty&0.0175&0.0199&0.0212&0.0247&0.0284&0.0365\\
         Toys&0.0097&0.0102&0.0168&0.0193&0.0320&0.0378\\
         ML-1M&0.0649&0.0720&0.0734&0.0786&0.0918&0.1032\\\bottomrule
    \end{tabular}
}
\end{table}

A simple method for ensemble modeling is to train multiple parallel networks independently and average their logits as the final output. We demonstrate its result in Table \ref{simpleensemble}. As we can see, the simple ensemble method indeed improves performance. But due to the absence of knowledge transfer between parallel networks, sometimes its performance margin can be trivial. Some works \cite{ensembleonthefly,CollaborativeLearning,DeepMutualLearning} have shown that an ensemble of networks can benefit from collaborative knowledge distillation, where each individual network acquires supplementary information from peer networks while sharing its own knowledge with others. Compared with independent training, such a collaborative paradigm can improve the performance of each individual network, where knowledge is effectively shared and transferred rather than learned alone. Therefore, it is necessary to design proper techniques for ensemble modeling methods that facilitate knowledge transfer between parallel networks.

Based on the above observations, we present \textbf{E}nsemble \textbf{M}odeling with contrastive \textbf{K}nowledge \textbf{D}istillation for sequential recommendation (EMKD). Our framework adopts multiple parallel networks as an ensemble of sequence encoders, and each parallel network is a bidirectional Transformer sequence encoder. To facilitate knowledge transfer between parallel networks, we propose a novel contrastive knowledge distillation approach, which distills knowledge from both the representation level and the logits level. At the representation level, two contrastive learning objectives are designed from both the intra-network perspective and cross-network perspective to distill representational knowledge. At the logits level, a collaborative knowledge distillation approach, which treats each parallel network as both the teacher and the student network, is designed to distill logits-level knowledge. For recommendation training, we introduce masked item prediction as the main task. Since attribute information incorporates rich contextual data that are useful for sequential recommendation \cite{FDSA,CIKM2020-S3Rec}, we design attribute prediction as an auxiliary task to fuse attribute information. 

To verify the effectiveness of our framework, we conduct extensive experiments on public benchmark datasets and show that our framework achieves significant performance improvement compared with the state-of-the-art methods. Furthermore, we show that our method for ensemble modeling is a generalized approach that can be adapted to other types of sequence encoders, such as RNN, CNN, and Transformer, to improve their performances. Since training efficiency can be a major concern for ensemble methods, we compare the training speed and convergence speed of our framework with other state-of-the-art methods and find out that our framework only shows a minor reduction in training efficiency. For these reasons, we think that it is worthwhile to adopt ensemble modeling methods in sequential recommendation. Our contributions can be summarized as follows:
\begin{itemize}
[leftmargin =  8pt,topsep=1pt]
    \item We propose a novel framework called Ensemble Modeling with Contrastive Knowledge Distillation for sequential recommendation (EMKD). To the best of our knowledge, this is the first work to apply the ensemble modeling to sequential recommendation.
    \item We propose a novel contrastive knowledge distillation approach that facilitates knowledge transfer and distills knowledge from both the representation level and the logits level.
    \item We conduct extensive experiments on public benchmark datasets and show that our framework can significantly outperform the state-of-the-art methods. Our framework can also be adapted to other sequential recommenders to improve their performances.
\end{itemize}

\section{RELATED WORK}
\subsection{Sequential Recommendation}
Traditional methods for sequential recommendation adopt the Markov Chains (MCs) to model item transitions, such as MDP \cite{MDP}, FPMC \cite{FPMC} and Fossil \cite{Fossil}. With the advancements in deep learning, RNN-based model \cite{srnn2016} and CNN-based model \cite{tang2018caser} have all been shown as effective sequence encoders. The recent success of Transformer \cite{vaswani2017transformer} also motivates the designing of Transformer-based sequence encoders. For example, SASRec \cite{kang18attentive} adopts unidirectional Transformer to automatically assign attention weights to different items, BERT4Rec \cite{Sun2019bert} adopts bidirectional Transformer with the cloze task \cite{Devlin2019BERT} to adapt for interactions that do not follow a rigid sequential order. Some works also leverage the attribute information about items in order to provide contextual data for sequential recommendation. For example, FDSA \cite{FDSA} adopts feature-level self-attention blocks to fuse attribute data into sequences, $\rm{S}^3$-Rec \cite{CIKM2020-S3Rec} proposes self-supervised objectives to model the correlations between items, sequences and attributes, MMInfoRec \cite{MMInfoRec} combines the attribute encoder with a memory module and a sampling strategy to capture long-term preferences. Different from these works that only focus on a single encoder, our framework trains an ensemble of networks with contrastive knowledge distillation.
\subsection{Ensemble Modeling and Knowledge Distillation}
Ensemble modeling \cite{dropout,snapshotensemble,ensembleonthefly,CollaborativeLearning,DeepMutualLearning,ensembleadversarialtraining,fastdnnensemble} has long been established as an effective approach for improving the accuracy and robustness of neural networks. One commonly adopted technique is Dropout \cite{dropout}, which creates a self-ensemble of networks by randomly dropping neurons at the training stage. As for the ensemble of multiple parallel networks, some works have recognized the necessity of facilitating knowledge when training multiple parallel networks, which is often implemented via knowledge distillation \cite{KnowledgeDistillation}. For example, DML \cite{DeepMutualLearning} proposes a collaborative learning strategy by distilling knowledge between parallel networks, ONE \cite{ensembleonthefly} proposes an on-the-fly native ensemble method for one-stage online distillation. However, these methods are all proposed for the compression of large-scale models on computer vision tasks, and how to design efficient and effective methods to facilitate knowledge transfer for ensemble sequential models remains to be explored. 
\subsection{Contrastive Learning}
Contrastive learning has achieved great success in computer vision \cite{contrastivepredictivecoding,mocov1,mocov2,mocov3,chen2020simple,SimSiam,ContrastiveRepresentationDistillation,fang2021seed,DistillingKnowledgeviaKnowledgeReview}, natural language processing \cite{gao2021simcse,yan-etal-2021-consert}, and recommender systems \cite{simgcl,CML,Xu2020Contrastive,DuoRec}. In sequential recommendation, contrastive learning has also been shown as an effective approach for alleviating data sparsity and learning better representations. For example, CL4SRec \cite{Xu2020Contrastive} first introduces contrastive learning into sequential recommendation by proposing three data augmentation operators, DuoRec \cite{DuoRec} performs contrastive learning from the model level to mitigate the representation degeneration problem, CML \cite{CML} combines contrastive learning with meta learning for personalized multi-behavior recommendation. Different from these works, our framework formulates contrastive learning as a way of contrastive representation distillation \cite{ContrastiveRepresentationDistillation} to facilitate knowledge transfer between the ensemble networks.
\begin{figure*}[htbp]
\centering
{\includegraphics[width=0.9\linewidth]{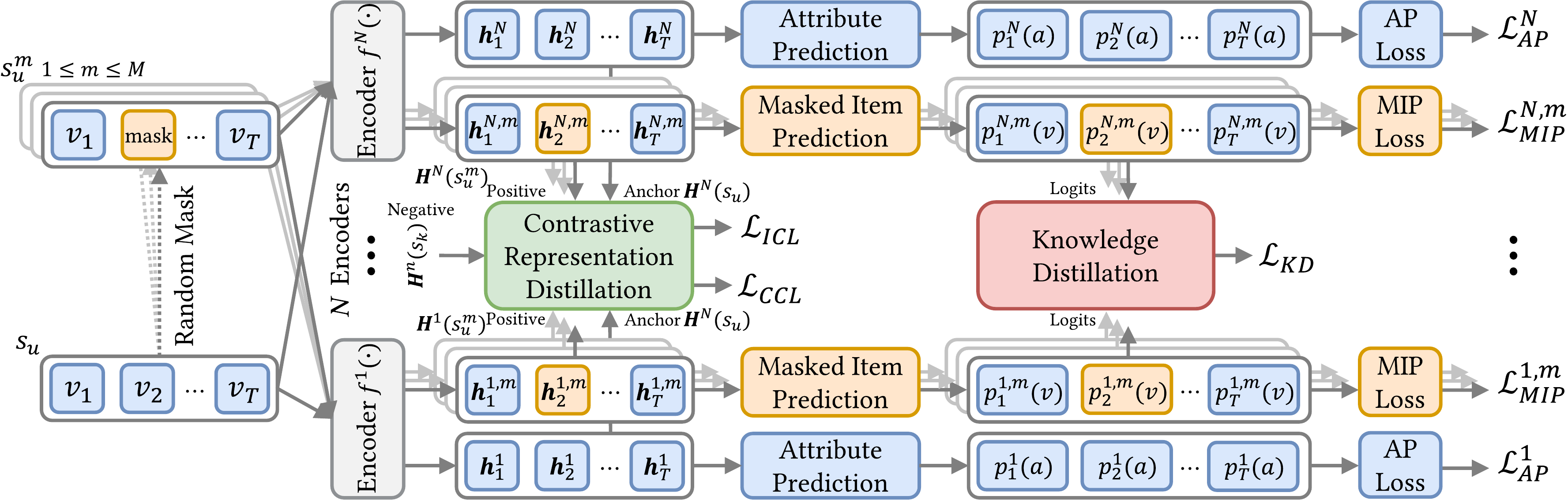}}

\caption{An overview of EMKD with $N$ parallel networks $f^{1}(\cdot),{\cdots},f^{N}(\cdot)$. For each original sequence $s_u$, we generate $M$ different masked sequences. The hidden representations of the original sequence $\boldsymbol{H}^{1}(s_u),\cdots,\boldsymbol{H}^{N}(s_u)$ serve as the anchor for contrastive representation distillation and are used for the attribute prediction task, while the hidden representations of the masked sequences $\boldsymbol{H}^{1}(s^{m}_{u}),\cdots, \boldsymbol{H}^{N}(s^{m}_{u})$ serve as positive samples for contrastive representation distillation and are used for the masked item prediction task. Negative samples $\boldsymbol{H}^{n}(s_{k})(1{\leq}n{\leq}N)$ for contrastive representation distillation are collected from the same batch. We compute the Kullback-Leibler divergence on the logits of the masked item prediction task between different networks for knowledge distillation.}\label{EMKD}
\end{figure*}
\section{PROPOSED FRAMEWORK}

In this section, we introduce the architecture of our proposed framework, Ensemble Modeling with contrastive Knowledge Distillation (EMKD). The architecture of EMKD is illustrated in Figure \ref{EMKD}. Our framework trains an ensemble of sequence encoders and averages the logits of these sequence encoders for inference. We design masked item prediction as the primary task for recommendation training and attribute prediction as the auxiliary task for attribute information fusion. To facilitate knowledge transfer between these parallel networks, we propose a novel approach for contrastive knowledge distillation, which performs contrastive representation distillation from the representation level and knowledge distillation from the logits level.
\subsection{Problem Statement}
In sequential recommendation, we denote $\mathcal{U}$ as a set of users, $\mathcal{V}$ as a set of items, and  $\mathcal{A}$ as a set of attributes. Each user $u\in\mathcal{U}$ is associated with an interaction sequence $s_u=[v_1,v_2,\cdots,v_t,\cdots,v_{T}\big]$ sorted in the chronological order, where $v_t\in \mathcal{V}$ denotes the item that user $u$ has interacted with at the $t$-th timestamp and $T$ indicates the sequence length. Each item $v_t$ has its own attribute set ${\alpha}_t=\{a_{1},a_{2},\cdots,a_{\lvert {\alpha}_t \rvert}\}\subset\mathcal{A}$ as a subset of all the attributes. The task of sequential recommendation is to predict the next item that user $u$ will probably interact with, and it can be formulated as generating the probability distribution over all candidate items for user $u$ at the time step $T+1$:\begin{displaymath}p(v_{T+1}=v|s_u)\end{displaymath}
\subsection{Ensemble Sequence Encoder}
A sequential recommendation model encodes essential sequence information via a sequence encoder, which takes user sequences as input and outputs hidden representations of sequences. In our framework, we adopt bidirectional Transformer as the base sequence encoder, which is able to model item correlations from both directions. The architecture of bidirectional Transformer consists of an embedding layer and a bidirectional Transformer module. For the input sequence $s_u=[v_1,v_2,\cdots,v_T]$, an item embedding matrix $\boldsymbol{V}\in {\mathbb{R}}^{{\lvert}\mathcal{V}{\rvert}\times d}$ and a positional embedding matrix ${\emph{\textbf{P}}}\in {\mathbb{R}}^{T\times d}$ and are combined together to represent items in the hidden space, where $T$ denotes the maximum sequence length of our model, $d$ is the hidden dimensionality. The computation of the embedding layer is formulated as follows:
\begin{equation}
\boldsymbol{E}(s_u)=[\boldsymbol{v}_{1}+\boldsymbol{p}_{1},\boldsymbol{v}_{2}+\boldsymbol{p}_{2},\cdots,\boldsymbol{v}_{T}+\boldsymbol{p}_{T}]
\end{equation}
We then feed $\boldsymbol{E}(s_u)$ into the bidirectional Transformer module $Trm$ and fetch the output $\boldsymbol{H}\in{\mathbb{R}^{T\times d}}$ as the hidden representations of the sequence. 
In general, we denote a sequence encoder $f(\cdot)$ that takes user sequence $s_u$ as the input and outputs the hidden representations $\boldsymbol{H}(s_u)$ of this sequence:
\begin{equation}
    \boldsymbol{H}(s_u)=[\boldsymbol{h}_{1},\boldsymbol{h}_{2},\cdots,\boldsymbol{h}_{T}]=f(s_u)=Trm(\boldsymbol{E}(s_u))
\end{equation}
As we have mentioned above, an ensemble of parallel networks can generate different predictions, which will be helpful for improving the overall accuracy rate. We adopt $N$ parallel networks as an ensemble of sequence encoders. Each parallel network $f^{n}(\cdot)$ is a bidirectional Transformer sequence encoder, where $1{\leq}n{\leq}N$. We initialize these parallel networks with different random seeds so that they will generate different prediction results.
\subsection{Multi-Task Learning}
We adopt a multi-task learning scheme to enhance sequential recommendation with contextual information. Specifically, we adopt masked item prediction as the main task for recommendation training. An attribute prediction task is also introduced as the auxiliary task to fuse attribute information.
\subsubsection{Masked Item Prediction}
Since bidirectional Transformers are typically trained with the masked item prediction task, we follow BERT4Rec \cite{Sun2019bert} and adopt masked item prediction as the main task, which requires the model to reconstruct the masked items. Specifically, for each user sequence $s_{u}$, we use different random seeds to generate $M$ different masked sequences. In each masked sequence $s^{m}_{u}$, a proportion $\rho$ of items ${\mathcal{I}}^{m}_{u}=(t^{m}_1,t^{m}_2,\cdots,t^{m}_{L})$ are randomly replaced with the mask token $\left[\rm{mask}\right]$. Here $L={\lceil}{\rho}*{T}{\rceil}$ and ${\mathcal{I}}^{m}_{u}$ denotes the indices of the masked items. The process of random item masking is defined as follows:
\begin{equation}
\begin{aligned}
    s^{m}_{u}&=\big[\widehat{v}_{1},\widehat{v}_{2},\cdots,\widehat{v}_{T}\big],1{\leq}m{\leq}M\\
\widehat{v}_t&=\left\{
\begin{array}{lc}
 v_t,& t\notin{\mathcal{I}}^{m}_{u} \\
 	\left[\rm{mask}\right],& t\in{\mathcal{I}}^{m}_{u}\\
\end{array}
\right.
\end{aligned}
\end{equation}
Each masked sequence $s^{m}_{u}$ is then fed into the ensemble networks to generate the corresponding hidden representations:
\begin{equation}
\boldsymbol{H}^{n}(s^{m}_u)=[\boldsymbol{h}^{n,m}_{1},\boldsymbol{h}^{n,m}_{2},\cdots,\boldsymbol{h}^{n,m}_{T}]=f^{n}(s^{m}_{u}),1{\leq}n{\leq}N,1{\leq}m{\leq}M
\end{equation}
A linear layer is adopted as the item classifier to convert the hidden representations into probability distribution over candidate items. Given the output of hidden representation ${\boldsymbol{h}}^{n,m}_{t}$ at position $t$, the computation is formulated as follows: 
\begin{equation}
\label{prob}
p^{n,m}_{t}(v)={\boldsymbol{h}}^{n,m}_{t}\boldsymbol{W}+\boldsymbol{b},1{\leq}t{\leq}T
\end{equation}
where $\boldsymbol{W}\in {\mathbb{R}}^{d\times {\lvert}\mathcal{V}{\rvert}}$ is the weight matrix and $\boldsymbol{b}\in \mathbb{R}^{\mathcal{\lvert V \rvert}}$ is the bias term. Finally, we calculate the cross-entropy loss for the $m$-th masked sequence from the $n$-th network:
\begin{equation}
\begin{aligned}
q^{n,m}_{t}(v_i)&=\frac{\mathrm{exp}(p^{n,m}_{t}(v_i))}{\sum_{j=1}^{|\mathcal{V}|}\mathrm{exp}(p^{n,m}_{t}(v_j))}\\
\mathcal{L}^{n,m}_{MIP}&=-\sum_{t\in{{\mathcal{I}}^{m}_{u}}}\sum_{i=1}^{|\mathcal{V}|}{y_{i}}\log{q^{n,m}_{t}(v_i)}
\end{aligned}
\end{equation}
where $y_{i}=1$ if $v_i=v_t$ (the predicted item $v_i$ is the ground-truth item $v_t$) else 0. Note that only the masked items within the sequence are considered when calculating the loss function for the masked item prediction task. Since we have $N$ parallel networks where each network is fed with $M$ different masked sequences, we sum up each $\mathcal{L}^{n,m}_{MIP}$ as the total loss function of the masked item prediction task:
\begin{equation}
    \mathcal{L}_{MIP}=\sum_{n=1}^{N}\sum_{m=1}^{M}\mathcal{L}^{n,m}_{MIP}
\end{equation}

\subsubsection{Attribute Prediction}
Attribute provides fine-grained features of items, which is a readily-available source of auxiliary information for sequential recommendation. Some works \cite{FDSA,CIKM2020-S3Rec,MMInfoRec} have shown that incorporating attribute information benefits the main task of sequential recommendation. Therefore, we propose an auxiliary attribute prediction task to incorporate attribute information. Different from previous works, we do not construct an explicit attribute embedding to model item-attribute correlations. Instead, we treat attribute prediction as a multi-label classification task that requires to the model to classify an item under the correct attribute(s). Specifically, we feed the original sequence $s_u$ into parallel networks to generate the hidden representations:
\begin{equation}
\boldsymbol{H}^{n}(s_u)=[\boldsymbol{h}^{n}_{1},\boldsymbol{h}^{n}_{2},\cdots,\boldsymbol{h}^{n}_{T}]=f^{n}(s_{u}),1{\leq}n{\leq}N
\end{equation}

We then adopt another linear layer with the sigmoid activation function as the attribute classifier to convert the hidden representation ${\boldsymbol{h}}^{n}_{t}$ at position $t$ into probability distribution over the attributes: 
\begin{equation}
p^{n}_{t}(a)=\sigma({\boldsymbol{h}}^{n}_{t}\boldsymbol{W}^{'}+\boldsymbol{b}^{'}), 1{\leq}t{\leq}T
\end{equation}
where $\boldsymbol{W}^{'}\in {\mathbb{R}}^{d{\times}{\lvert}\mathcal{A}{\rvert}}$ is the weight matrix, $\boldsymbol{b}^{'}\in \mathbb{R}^{{\lvert}\mathcal{A}{\rvert}}$ is the bias term, and $\sigma$ is the sigmoid function. Given that an item may be associated with multiple attributes, we compute the binary cross-entropy loss function for the attribute prediction task from the $n$-th network:
\begin{equation}
\mathcal{L}^{n}_{AP}=-\sum^{T}_{t=1}\sum_{i=1}^{{\lvert}\mathcal{A}{\rvert}}\left[{y_{i}}\log{p^{n}_{t}(a_i)}+{(1-y_{i})}{\log( 1-{p^{n}_{t}(a_i)}})\right]
\end{equation}
where $y_{i}=1$ if $a_i\in{\alpha}_t$ (the predicted attribute $a_i$ is a correct attribute in the associated attribute set ${\alpha}_t$ of item $v_t$) else 0. Each $\mathcal{L}^{n}_{AP}$ is then summed up as the total loss function of the attribute prediction task:
\begin{equation}
    \mathcal{L}_{AP}=\sum_{n=1}^{N}\mathcal{L}^{n}_{AP}
\end{equation}

\subsection{Contrastive Knowledge Distillation}
To transfer knowledge between parallel networks, we design a contrastive knowledge distillation approach that transfers both the representational knowledge and the logits-level knowledge. 
\subsubsection{Contrastive Representation Distillation.}
Contrastive learning aims to align the positive sample (i.e., a data-augmented sequence) with its anchor (i.e., an original sequence) and push negative samples apart from the anchor. In our framework, the hidden representation ${\boldsymbol{H}}^{n}(s_u)$ from the attribute prediction task can be viewed as an anchor representation because it is obtained from the original user sequence $s_u$. Besides, the other hidden representations ${\boldsymbol{H}}^{n}(s^{m}_{u})$ obtained from the masked item prediction task can be viewed as positive samples, because they are obtained from masked user sequences $s^{m}_{u}$, which is a form of data augmentation. Similar to the in-batch negative sampling strategy in \cite{chen2020simple,Xu2020Contrastive,DuoRec}, the other user sequences $\{s_k\}^{\mathcal{B}}_{k=1,k{\neq}u}$ from the same batch can be viewed as negative samples, where $\mathcal{B}$ denotes the batch size.

Our method for contrastive representation distillation contains two objectives. First, an Intra-network Contrastive Learning (ICL) objective is designed to contrast the hidden representations of different masked sequences outputted by the same network. The loss function of ICL for the $n$-th network is formulated as follows:
\begin{equation}
    \label{ICL}
    \mathcal{L}^{n}_{ICL}=-\sum^{M}_{m=1}\log{\frac{\mathrm{exp}(g({\boldsymbol{H}}^{n}(s_u),{\boldsymbol{H}}^{n}(s
    ^{m}_{u}))/\tau)}{\sum^{\mathcal{B}}_{k=1,k{\neq}u}\mathrm{exp}(g({\boldsymbol{H}}^{n}(s_u),{\boldsymbol{H}}^{n}(s_{k}))/\tau)}}
\end{equation}
where $g(\cdot)$ is the cosine similarity function, $\tau$ is a temperature hyper-parameter, $M$ is the number of the masked sequences. Each $\mathcal{L}^{n}_{ICL}$ is then summed up as the total loss function for ICL:
\begin{equation}
    \mathcal{L}_{ICL}=\sum^{N}_{n=1}\mathcal{L}^{n}_{ICL}
\end{equation}

ICL alone cannot effectively transfer knowledge between the parallel networks, because the representations from different networks are not contrasted against each other. To transfer representational knowledge between the parallel networks, a Cross-network Contrastive Learning (CCL) objective is introduced to contrast the hidden representations of different masked sequences outputted by different networks. Similar to Eq. \ref{ICL}, the loss function of CCL between the $x$-th network and the $y$-th network is formulated as follows:
\begin{equation}
    \label{CCL}
    \mathcal{L}^{x,y}_{CCL}=-\sum^{M}_{m=1}\log{\frac{\mathrm{exp}(g({\boldsymbol{H}}^{x}(s_u),{\boldsymbol{H}}^{y}(s^{m}_u))/\tau)}{\sum^{\mathcal{B}}_{k=1,k{\neq}u}\mathrm{exp}(g({\boldsymbol{H}}^{x}(s_u),{\boldsymbol{H}}^{y}(s_{k}))/\tau)}}
\end{equation}
where $x,y\in[1,N]$ are the indices of the networks. Since we have $N$ parallel networks, permutations of $\mathcal{L}^{x,y}_{CCL}$ are summed up as the total loss function for CCL:
\begin{equation}
    \mathcal{L}_{CCL}=\sum^{N}_{x=1}\sum^{N}_{y=1,y{\neq}x}\mathcal{L}^{x,y}_{CCL}
\end{equation}
\subsubsection{Logits-Level Knowledge Distillation.}
The task of sequential recommendation takes the logits $p^{n,m}_{t}(v)$ as the final prediction result. Although representational knowledge is distilled through contrastive learning, we still need Knowledge Distillation (KD) from the logits level so as to transfer knowledge more related with the sequential recommendation task. Following \cite{KnowledgeDistillation,ensembleonthefly} for knowledge distillation, we minimize the Kullback-Leibler divergence between the teacher logits $p^{x,m}_{t}(v)$ and the student logits $p^{y,m}_{t}(v)$. Specifically, we first compute the softmax probability distribution at temperature $\tau$:
\begin{equation}
z^{x,m}_{t}(v_i)=\frac{\mathrm{exp}(p^{x,m}_{t}(v_i)/\tau)}{\sum_{j=1}^{|\mathcal{V}|}\mathrm{exp}(p^{x,m}_{t}(v_j)/\tau)},z^{y,m}_{t}(v_i)=\frac{\mathrm{exp}(p^{y,m}_{t}(v_i)/\tau)}{\sum_{j=1}^{|\mathcal{V}|}\mathrm{exp}(p^{y,m}_{t}(v_j)/\tau)}
\end{equation}
The computation of the Kullback-Leibler divergence between the $x$-th network and the $y$-th network can then be formulated as follows:
\begin{equation}
\begin{aligned}
    \mathcal{L}^{x,y}_{KD}=\sum^{M}_{m=1}\sum_{t\in{{\mathcal{I}}^{m}_{u}}}\sum_{i=1}^{|\mathcal{V}|}z^{x,m}_{t}(v_i)\log\frac{{z}^{x,m}_{t}(v_i)}{{z}^{y,m}_{t}(v_i)}
\end{aligned}
\end{equation}
where the probability distributions of the $M$ masked sequences are all considered for knowledge distillation.
Note that when $z^{x,m}_{t}(v_i)$ is treated as the teacher logits, we cut off its gradient for stability. 

The logits-level knowledge is distilled in a collaborative style, where each network can simultaneously be the student who acquires knowledge from other peer networks and the teacher who provides knowledge for other peer networks. In this way, the distilled knowledge can be effectively shared and transferred between all parallel networks. Therefore, we sum up the permutations of $\mathcal{L}_{KD}^{x,y}$ for all the $N$ parallel networks to perform a collaborative paradigm of knowledge distillation:
\begin{equation}
    \mathcal{L}_{KD}=\sum^{N}_{x=1}\sum^{N}_{y=1,y{\neq}x}\mathcal{L}^{x,y}_{KD}
\end{equation}

\subsection{Training and Inference}
\subsubsection{Overall Training Objective.} We sum up the loss function for the masked item prediction task, the attribute prediction task, the ICL objective, the CCL objective, and the KD objective as the total loss function to jointly optimize our framework:
\begin{equation}
    \mathcal{L}_{EMKD}=\mathcal{L}_{MIP}+\mathcal{L}_{AP}+\lambda(\mathcal{L}_{ICL}+\mathcal{L}_{CCL})+\mu{\mathcal{L}}_{KD}
\end{equation}
where $\lambda,\mu$ are weight hyper-parameters.
\subsubsection{Inference}
To infer the next item for user $u$, we append the mask token to the end of the sequence $s_{u}$ to obtain $s_{u^{'}}=\left[v_{2},v_{3}\cdots,v_{T},[\mathrm{mask}]\right]$\footnote{We drop the first item $v_{1}$ due to the restriction of maximum sequence length.}. The hidden representations corresponding to the mask token are converted into probability distributions by the item classifier defined in Equation \ref{prob}. Predictions from all the parallel networks are averaged as the final prediction result:
\begin{equation}
\begin{aligned}
\boldsymbol{H}^{n}(s_{u^{'}})&=[\boldsymbol{h}^{n}_{2},\boldsymbol{h}^{n}_{3},\cdots,\boldsymbol{h}^{n}_{T},\boldsymbol{h}^{n}_{\rm{mask}}]=f^{n}(s_{u^{'}})\\
p(v)&=\frac{1}{N}\sum^{N}_{n=1}({\boldsymbol{h}}^{n}_{\rm{mask}}\boldsymbol{W}+\boldsymbol{b})
\end{aligned}
\end{equation}
\subsection{Discussion}
One possible reason why ensemble methods are superior for sequential recommendation is that it can handle behavior uncertainty. Users' sequential behaviors are uncertain rather than deterministic, and users' preferences may shift under different circumstances \cite{STOSA}. For example, a user favours movies with Action and Comedy genres. However, there are many candidate movies that fall under these genres, and the user may choose different movies under different situations. In such cases, the diversity of prediction results brought by ensemble methods can consider all these situations and generate a more accurate recommendation result.

Another possible reason is that ensemble methods can mitigate the adverse effects of noisy data. The datasets for sequential recommendation are usually sparse and noisy. In such cases, the knowledge transfer mechanisms between the ensemble networks can filter out the incorrect predictions introduced by noisy labels and transfer the useful knowledge to the whole network \cite{CoTeaching}, thus improving the overall accuracy rate and model robustness.
\section{EXPERIMENT}\label{experiment}
In this section, we present the details of our experiments and answer the following research questions (\textbf{RQs}):
\begin{itemize}
[leftmargin =  8pt,topsep=1pt]
\item\textbf{RQ1}: How does EMKD perform comparing with other state-of-the-art methods for sequential recommendation? 
\item\textbf{RQ2}: What are the influences of different hyper-parameters in EMKD?
\item\textbf{RQ3}: What is the effectiveness of different components in EMKD?
\item\textbf{RQ4}: Can we adapt EMKD to other sequential recommenders to improve their performances?
\item\textbf{RQ5}: Compared with other state-of-the-art methods, will ensemble modeling significantly reduce training efficiency?
\item\textbf{RQ6}: Does the good performance of EMKD result from the proposed knowledge transfer mechanisms between the ensemble networks or does it simply come from the increase in the parameter size?
\end{itemize}
\subsection{Settings}
\subsubsection{Dataset.}
We conduct experiments on three public benchmark datasets collected from two platforms. The Amazon dataset \cite{mcauley15image} contains users' reviews on varying categories of products. We select two subcategories, \textbf{Beauty} and \textbf{Toys}, and use the fine-grained categories and the brands of the products as attributes. Another dataset MovieLens-1M (\textbf{ML-1M}) \cite{movielens} contains users' ratings on movies, and we use the genres of the movies as attributes.

Following \cite{kang18attentive,Sun2019bert,CIKM2020-S3Rec,Xu2020Contrastive,DuoRec}, we treat all interactions as implicit feedbacks. To construct user sequences, we remove duplicated interactions and sort each user's interactions by their timestamps. Users related with less than 5 interactions and items related with less than 5 users are filtered out from the dataset. The processed dataset statistics are presented in Table \ref{datasetstatistics}.
\begin{table}
    \centering
    \caption{Dataset statistics after preprocessing.}
    \label{datasetstatistics}
  \setlength{\tabcolsep}{1pt}{
    \begin{tabular}{lccc}
    \toprule
         Datasets&Beauty&Toys&ML-1M\\\midrule
         \#users&22,363&19,412&6,040\\
         \#items&12,101&11,924&3,953\\
         \#actions&198,502&167,597&1,000,209\\
         avg. actions/user&8.9&8.6&163.5\\
         avg. actions/item&16.4&14.1&253.0\\
         sparsity&99.93\%&99.93\%&95.81\%\\
         \#attributes&1,221&1,027&18\\
         avg. attributes/item&5.1&4.3&1.7
 \\\bottomrule
    \end{tabular}
}
\end{table}
\subsubsection{Metrics.}
 We choose top-$K$ Hit Ratio (HR@$K$) and Normalized Discounted Cumulative Gain (NDCG@$K$) with $K\in\{5,10\}$ as metrics to evaluate the performance. The \textit{leave-one-out} evaluation strategy is adopted, holding out the last item for test, the second-to-last item for validation, and the rest for training. Since sampled metrics might lead to unfair comparisons \cite{krichene2020sampled}, we rank the prediction results on the whole item set without sampling.
\subsubsection{Baselines.}
We include baseline methods from three groups for comparison:
\begin{itemize}
[leftmargin =  8pt,topsep=1pt]
    \item\textbf{General sequential methods} utilize a sequence encoder to generate the hidden representations of users and items. For example, GRU4Rec \cite{srnn2016} adopts RNN as the sequence encoder, Caser \cite{tang2018caser} adopts CNN as the sequence encoder, SASRec \cite{kang18attentive} adopts unidirectional Transformer as the sequence encoder, BERT4Rec \cite{Sun2019bert} adopts bidirectional Transformer as the sequence encoder.
    \item\textbf{Attribute-aware sequential methods} fuse attribute information into sequential recommendation. For example, FDSA \cite{FDSA} applies self-attention blocks to capture transition patterns of items and attributes, $\rm{S}^3$-Rec \cite{CIKM2020-S3Rec} proposes self-supervised objectives to model the correlations between items, sequences and attributes, MMInfoRec \cite{MMInfoRec} augments the attribute encoder with a memory module and a multi-instance sampling strategy.
    \item\textbf{Contrastive sequential methods} design auxiliary objectives for contrastive learning based on general sequential methods. For example, CL4SRec \cite{Xu2020Contrastive} proposes data augmentation strategies for contrastive learning in sequential recommendation, DuoRec \cite{DuoRec} proposes both supervised and unsupervised sampling strategies for contrastive learning in sequential recommendation.
\end{itemize}
\subsubsection{Implementation.}
The codes for GRU4Rec, Caser, SASRec, BERT4Rec, $\rm{S}^3$-Rec, MMInfoRec, and DuoRec are provided by their authors. We implement FDSA and CL4SRec in PyTorch \cite{pytorch}. We follow the instructions from the original papers to set and tune the hyper-parameters. 

We also implement our framework in PyTorch \cite{pytorch}. We set the number of Transformer blocks and attention heads as 2, batch size as 256. Following BERT \cite{Devlin2019BERT} we set the mask proportion $\rho$ as 0.15. For other hyper-parameters, we tune $\lambda$ within $[0.01,1]$, $\mu$ within $[0.01,1]$, $\tau$ within $[0.1,10]$, the number of masked sequences $M$ within $[1,8]$. The default setting for the number of parallel networks $N$ is 3, and variants are studied in Section \ref{ablationstudy}. The default setting for the hidden dimensionality $d$ is set as 256, and model robustness w.r.t different hidden dimensionality sizes is studied in Section \ref{hiddendim}. We use an Adam optimizer \cite{Adam} with an initial learning rate 0.001, and the learning rate decays exponentially after every 100 epochs. We train our model for 250 epochs and select the checkpoint with the best validation result for test.

\subsection{Overall Performance Comparison (RQ1)}\label{overallperformance}
\begin{table*}

  \centering
     \caption{Overall performance of different methods for sequential recommendation. The best score and the second-best score in each row are bolded and underlined, respectively. The last column indicates improvements over the best baseline method.}
  \setlength{\tabcolsep}{4pt}{

    \begin{tabular}{c|l|cccc|ccc|cc|c|c}
\toprule
Dataset&Metric&GRU4Rec&Caser&SASRec &BERT4Rec&FDSA&$\rm{S^3}$-Rec&MMInfoRec&CL4SRec&DuoRec&EMKD&Improv. \\
\midrule   \multirow{4}[0]{*}{Beauty} 
&HR@5&0.0206&0.0254&0.0371&0.0364&0.0317&0.0382&0.0527&0.0396&\underline{0.0559}&\textbf{0.0702}& 25.58\%\\
&HR@10  &0.0332&0.0436&0.0592&0.0583&0.0496&0.0634 &0.0739&0.0630&\underline{0.0867}&\textbf{0.0995}& 14.76\%\\
&NDCG@5 &0.0139&0.0154&0.0233&0.0228&0.0184&0.0244&\underline{0.0378}&0.0232&0.0331&\textbf{0.0500}& 32.28\%\\
&NDCG@10&0.0175&0.0212&0.0284&0.0307&0.0268&0.0335&\underline{0.0445}&0.0307&0.0430&\textbf{0.0594}& 33.48\%\\
\midrule  \multirow{4}[0]{*}{Toys} 
&HR@5 &0.0121&0.0205& 0.0429  & 0.0371 & 0.0269&0.0440&\underline{0.0579}& 0.0503 & 0.0539& \textbf{0.0745}& 28.67\%\\
&HR@10&0.0184&0.0333&0.0652&0.0524&0.0483&0.0705&\underline{0.0818}&0.0736&0.0744& \textbf{0.1016}  & 24.21\%\\
&NDCG@5&0.0077&0.0125&0.0248&0.0259&0.0227&0.0286&\underline{0.0408}&0.0264&0.0340&\textbf{0.0534} & 30.88\%\\
&NDCG@10&0.0097&0.0168&0.0320&0.0309&0.0281&0.0369&\underline{0.0484}&0.0339&0.0406&\textbf{0.0622} & 28.51\%\\
\midrule    
  \multirow{4}[0]{*}{ML-1M} 
&HR@5&0.0806&0.0912&0.1078&0.1308&0.0953&0.1128&0.1454&0.1142&\underline{0.1930}&\textbf{0.2315}&19.95\%\\
&HR@10&0.1344&0.1442& 0.1810 & 0.2219 & 0.1645 & 0.1969&0.2248&0.1815&\underline{0.2865}& \textbf{0.3239}&13.05\%\\
&NDCG@5&0.0475&0.0565&0.0681&0.0804&0.0597&0.0668&0.0856&0.0705&\underline{0.1327}&\textbf{0.1616}&21.78\%\\
&NDCG@10&0.0649&0.0734&0.0918&0.1097&0.0864&0.0950&0.1203&0.0920&\underline{0.1586}&\textbf{0.1915}&20.74\%\\
\bottomrule\end{tabular}}
  \label{result}
\end{table*}
Table \ref{result} presents the overall performance of EMKD and baseline methods. Based on the experiment results, we can see:
\begin{itemize}
[leftmargin =  8pt,topsep=1pt]
\item EMKD outperforms all baseline methods on both sparse and dense datasets with the relative performance improvements ranging from 13.05\% to 33.48\%. We attribute the performance improvement to these factors: (1) ensemble modeling uses multiple parallel networks to yield diverse predictions, and combining a diversity of prediction results is more accurate than the prediction of a single network; (2) contrastive knowledge distillation facilitates knowledge transfer between parallel networks, and each individual network benefits from this collaborative learning paradigm; (3) the attribute information provides rich contextual data and can benefit the main task of sequential recommendation.

\item Among general sequential methods, Transformer-based sequence encoders (e.g., SASRec, BERT4Rec) outperform RNN-based (e.g., GRU4Rec) or CNN-based (e.g., Caser) sequence encoders. This suggests that the self-attention mechanism can effectively model sequential patterns. Moreover, attribute-aware sequential methods outperform general sequential methods owing to the incorporation of attribute information. Besides, contrastive sequential methods show performance improvements compared with general sequential methods. This is probably because contrastive
learning serves as a regularization objective that can alleviate the
data sparsity issue and improve model robustness.
\end{itemize}
\subsection{Hyper-parameter Sensitivity (RQ2)}\label{hyperparametersensitivity}
In this section, we study the influences of important hyper-parameters, including the number of masked sequences $M$, temperature $\tau$, weight hyper-parameters $\lambda$ and $\mu$, hidden dimensionality $d$. To control variables we only change one hyper-parameter at one time while keeping the other hyper-parameters optimal.

\begin{figure*}
\centering

\subcaptionbox{Number of Masked Seq.\label{positivesamples}}
{\includegraphics[width=.19\linewidth]{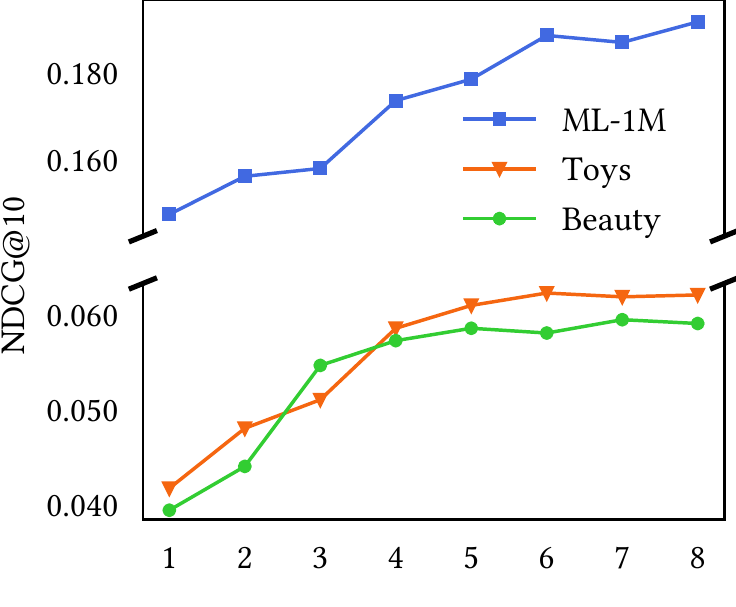}}
\subcaptionbox{Temperature $\tau$\label{temperature}}
{\includegraphics[width=.19\linewidth]{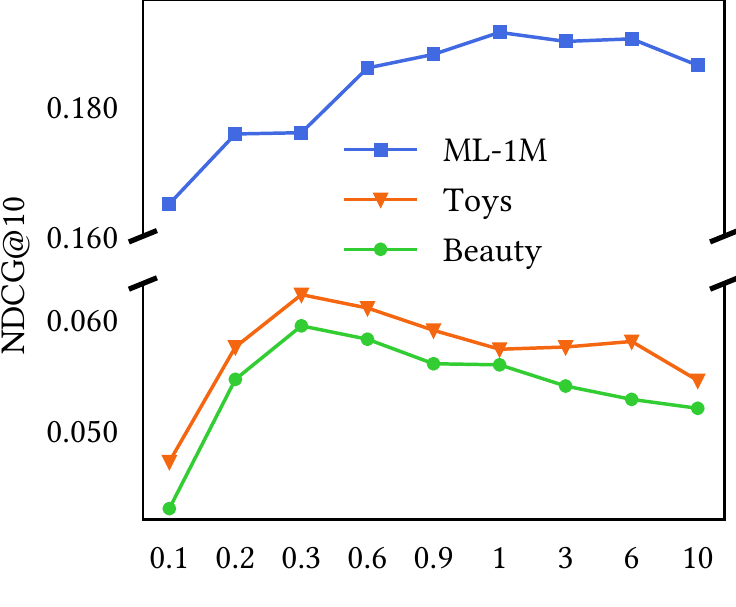}}
\subcaptionbox{Hyper-parameter $\lambda$\label{lambda}}
{\includegraphics[width=.19\linewidth]{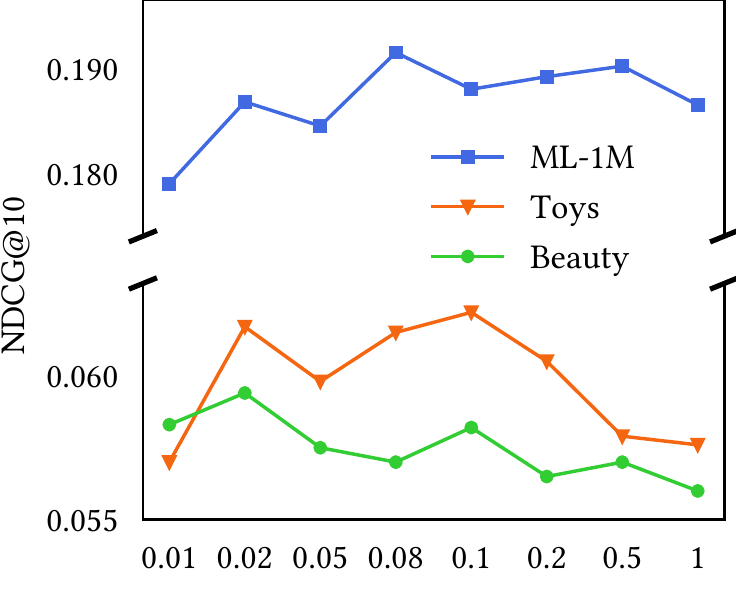}}
\subcaptionbox{Hyper-parameter $\mu$\label{mu}}
{\includegraphics[width=.19\linewidth]{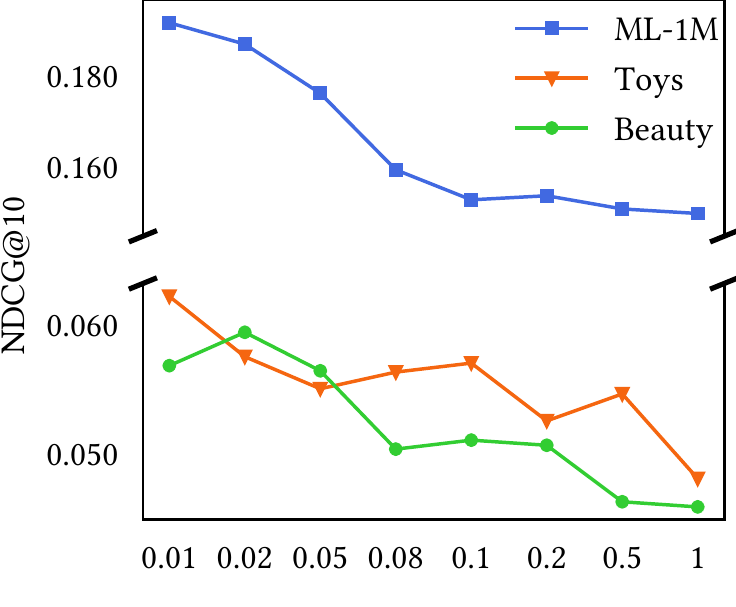}}
\subcaptionbox{Hidden Dimensionality \label{d}}
{\includegraphics[width=.19\linewidth]{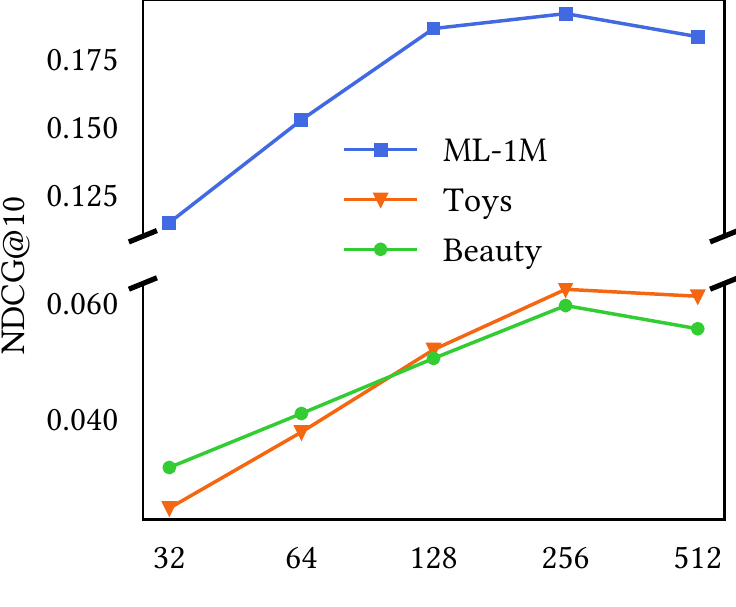}}

\caption{Performance (NDCG@10) comparison w.r.t different hyper-parameters on three datasets.}\label{hyperparameter}
\end{figure*}
\subsubsection{Number of Masked Sequences.} $M$ regulates how many masked sequences are available for training. From Figure \ref{positivesamples}, we can see that the performance increases as more masked sequences are available. This is because masking different items in the same sequence brings various semantic patterns, and reconstructing all these items will help the model gain more knowledge. However, such performance gain also shows an upper limit for multiple masked sequences. With sufficient masked sequences, the performance hardly improves even if we further increase $M$.
\subsubsection{Temperature.} Temperature $\tau$ regulates the hardness of the probability distributions for contrastive representation distillation and knowledge distillation. From Figure \ref{temperature}, we can see that an appropriate choice of $\tau$ should be neither too large nor too small. This is because a smaller $\tau$ produces harder probability distributions that makes the model intolerant to semantically similar samples, while a larger $\tau$ produces softer probability distributions that makes the model insensitive to semantic differences \cite{wang2020alignmentuniformity,wang2021understanding}.
\subsubsection{Weight Hyper-parameters.}$\lambda$ and $\mu$ are two weight hyper-parameters that control the strength of contrastive representation distillation and knowledge distillation. From Figure \ref{lambda} and Figure \ref{mu}, we can see that a proper choice of weight hyper-parameters can significantly improve the performance, while selecting weight hyper-parameters that are either too large or too small will undermine the performance.

\subsubsection{Hidden Dimensionality $d$.}\label{hiddendim} Figure \ref{d} reports the performances w.r.t different hidden dimensionality $d$. We can see that the performance steadily increases as the hidden dimensionality becomes larger, and the best performance is reached when $d=256$ (default setting). However, the performance slightly drops when $d=512$, which is probably due to overfitting.

\subsection{Ablation Study (RQ3)}
\begin{table}
    \centering
    \small
    \caption{Ablation study (NDCG@10) on three datasets. Bold score indicates the performance under the default setting. $\uparrow$ indicates the performance better than the default setting.}
    
  \setlength{\tabcolsep}{5pt}{
    \begin{tabular}{l|ccc}
    \toprule
    
    \multirow{2}{*}{Architecture} & \multicolumn{3}{c}{Dataset} \\& Beauty&Toys&ML-1M\\\midrule
         (1) EMKD($\times$3)&\textbf{0.0594}&\textbf{0.0622}&\textbf{0.1915}\\\midrule
         (2) Remove ICL&0.0529&0.0545&0.1679\\
         (3) Remove CCL&0.0552&0.0560&0.1807\\
         (4) Remove KD&0.0537&0.0571&0.1758\\
         (5) Independent Training&0.0452&0.0484&0.1476\\
         \midrule
         (6) Single Encoder & 0.0363 & 0.0375 & 0.1183\\
         (7) EMKD($\times$2)&0.0536&0.0568&0.1792\\
         (8) EMKD($\times$4)&0.0591&0.0629$\uparrow$&0.1930$\uparrow$\\\midrule
         (9) Remove AP&0.0578&0.0609&0.1831
 \\\bottomrule
    \end{tabular}
}
\label{ablationstudytable}
\end{table}
In this section, we conduct an ablation study to investigate the effectiveness of the key components in our framework. Table \ref{ablationstudytable} presents the performance of EMKD under the default setting and its variants. Based on the experiment results, we can see that:
\begin{itemize}
[leftmargin =  8pt,topsep=1pt]

\item\textbf{Ensemble networks outperform a single network.} We can see that (6) shows the worst performance compared with other ensemble methods. This observation verifies our claim that ensemble modeling is more powerful than a single network. 
\item\textbf{Ensemble networks learn better with effective knowledge transfer.} Our method for contrastive knowledge distillation contains three objectives---ICL, CCL and KD. From (2)-(4) we can see that each objective plays a crucial role in facilitating knowledge transfer between parallel networks, and removing any objective leads to performance decrease. Compared with (5), we can see that ensemble networks with knowledge transfer methods outperform the simple ensemble of independently-trained networks.
\item \textbf{Training EMKD with 3 parallel networks is the optimal setting.} Increasing the number of parallel networks can improve performance because training more parallel networks can increase the diversity of prediction results. However, we cannot train an infinite number of parallel networks due to the limitation of computational cost. From (1)(6)(7)(8) we can see that training EMKD with 3 parallel networks almost reaches the best performance. Although (8) shows a slight improvement compared with (1), training EMKD with 4 parallel networks is too costly and also increases the risk of overfitting.
\item \textbf{Sequential recommendation benefits from the auxiliary attribute information.} From (9) we can see that removing the auxiliary task of attribute prediction will hurt performance. This observation is also consistent with previous works \cite{FDSA,CIKM2020-S3Rec,MMInfoRec} that utilizing attribute information can enhance the performance of sequential recommendation models. 
\end{itemize}
\label{ablationstudy}

\begin{table}[htbp]
    \centering
    \footnotesize
    \caption{Performance comparison (NDCG@10) of models with different parameter sizes on three datasets. $*$ indicates the default setting for each model.}
    
  \setlength{\tabcolsep}{1.35pt}{
    \begin{tabular}{lcccccc}
    \toprule &\multicolumn{2}{c}{Beauty}&\multicolumn{2}{c}{Toys}&\multicolumn{2}{c}{ML-1M}\\\cmidrule(lr){2-3}\cmidrule(lr){4-5}\cmidrule(lr){6-7}
    Architecture&Params.&NDCG@10&Params.&NDCG@10&Params.&NDCG@10\\
    \midrule
         SASRec-2 Layers$^{*}$&4.69M&0.0284&4.65M&0.0320&2.51M&0.0918\\
         SASRec-4 Layers&6.27M&0.0301&6.23M&0.0313&4.09M&0.0896\\
         SASRec-6 Layers&7.85M&0.0298&7.80M&0.0332&5.67M&0.0857\\
         SASRec-8 Layers&9.43M&0.0279&9.38M&0.0305&7.24M&0.0932\\
         SASRec-10 Layers&11.01M&0.0282&10.96M&0.0310&8.82M&0.0881\\
         \midrule
         BERT4Rec-2 Layers$^{*}$& 7.80M & 0.0307&7.71M&0.0309&3.53M&0.1097\\
         BERT4Rec-4 Layers& 9.38M &0.0328&9.29M&0.0312&5.11M&0.1113\\
         BERT4Rec-6 Layers& 10.96M&0.0332&10.87M&0.0306&6.69M&0.1100\\
         BERT4Rec-8 Layers&12.54M&0.0310&12.45M&0.0298&8.27M&0.1093\\
         BERT4Rec-10 Layers&14.12M&0.0319&14.03M&0.0293&9.85M&0.1099\\
         \midrule
         
         EMKD($\times$2)&9.36M&0.0536&9.28M&0.0568&5.08M&0.1792\\
         EMKD($\times$3)$^{*}$&14.05M&0.0594&13.91M&0.0622&7.62M&0.1915
 \\\bottomrule
    \end{tabular}}
\label{parameterscalingtable}
\end{table}
\subsection{Adaptation to Other Models (RQ4)}
EMKD trains an ensemble of bidirectional Transformers as sequence encoders, which can be viewed as an ensemble of BERT4Rec models. Since ensemble modeling and knowledge transfer is a model-agnostic approach, it can also be adapted to other sequence encoders. In this section, we apply EMKD to the ensemble of various base sequence encoders with a slight modification. Different from the Masked Language Modeling (MLM) style in bidirectional Transformers, other sequence encoders (e.g., GRU4Rec, Caser and SASRec) are autoregressive models that predict the next item based on previous items. Therefore, we need to switch the masked item prediction task to the next item prediction task. The attribute prediction task is also modified so that the model will predict the associate attributes at the next position.

\begin{figure}
\centering
\subcaptionbox{Beauty\label{adaptation_beauty}}
{\includegraphics[ width=.4\linewidth]{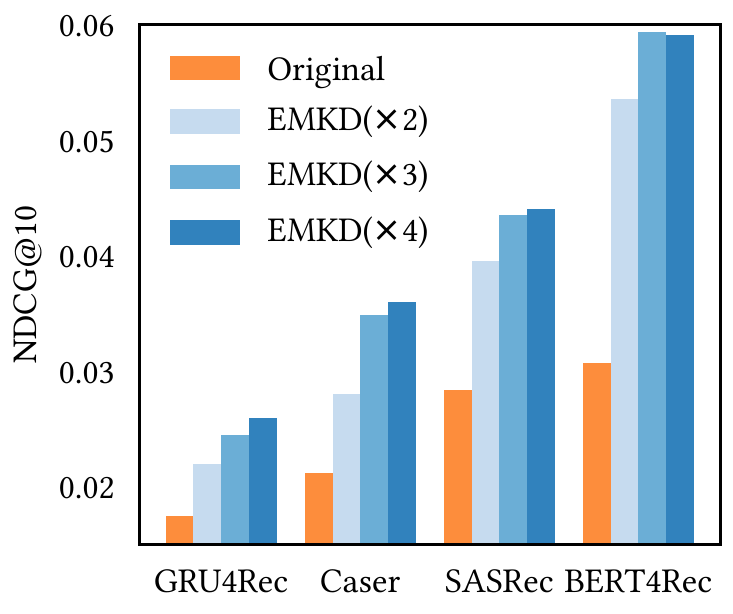}}
\subcaptionbox{ML-1M\label{adaptation_ML_1M}}
{\includegraphics[ width=.4\linewidth]{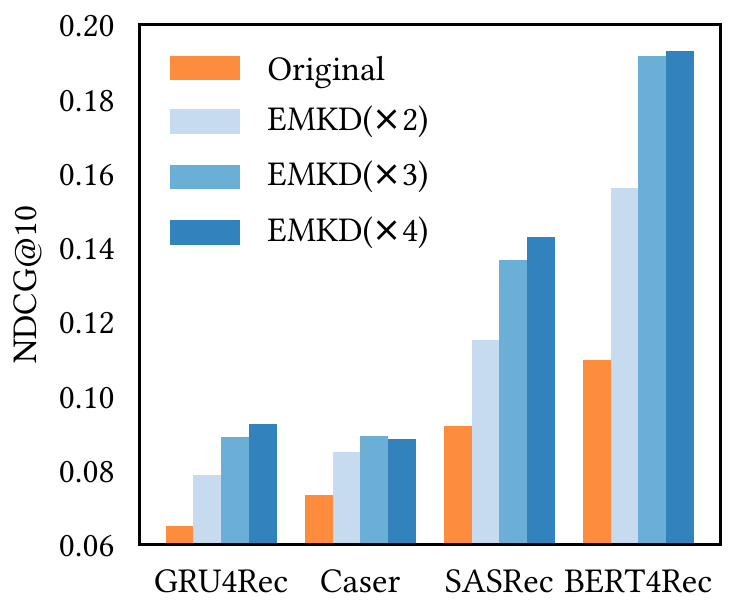}}
\caption{Performance comparison (NDCG@10) of different models enhanced by EMKD on Beauty and ML-1M datasets. We design three variants for each group of base sequence encoder with 2,3,4 parallel networks respectively.}
\label{adaptation}
\end{figure}

Figure \ref{adaptation} presents the performances of different models enhanced by EMKD on Beauty and ML-1M datasets. We can see that all the base sequence encoders enhanced by our approach achieve better performance. This shows that ensemble modeling with contrastive knowledge distillation is a generalized approach that can be utilized to improve the performance of sequential recommenders.

\subsection{Training Efficiency (RQ5)} 
\begin{figure}
\centering
\subcaptionbox{Beauty\label{efficienty_beauty}}
{\includegraphics[ width=.4\linewidth]{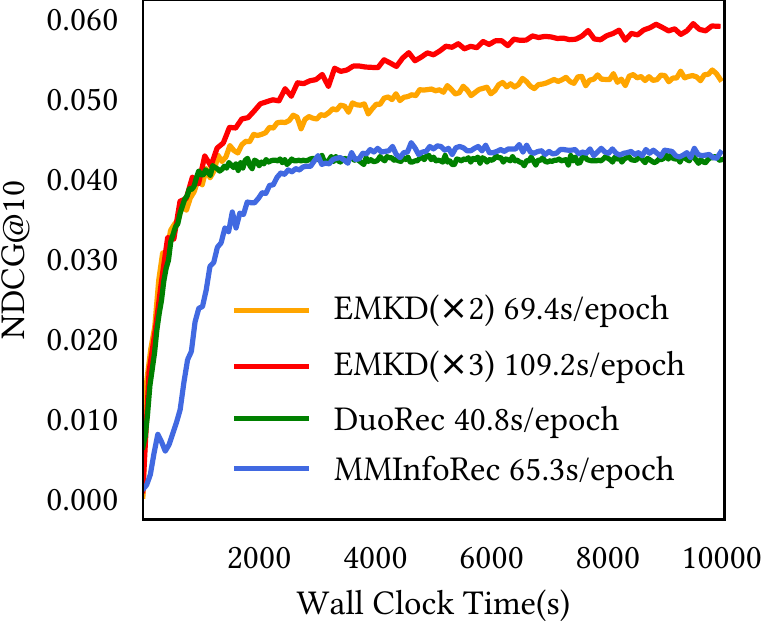}}
\subcaptionbox{Toys\label{efficienty_toys}}
{\includegraphics[ width=.4\linewidth]{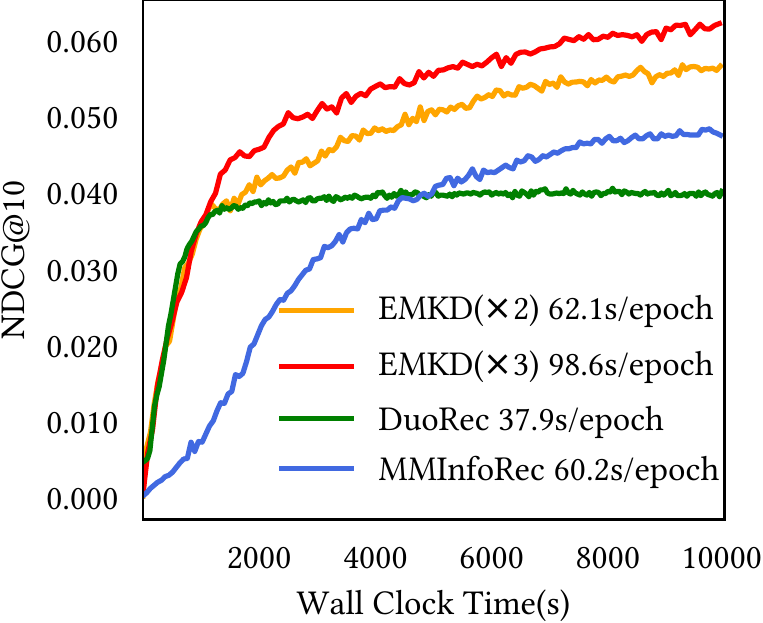}}
\caption{Training efficiency (NDCG@10) on Beauty and Toys datasets. The training speed of EMKD is slightly lower than MMInfoRec, while the convergence speed of EMKD is comparable with DuoRec.}
\label{trainingefficiency_figure}
\end{figure}
A major concern for ensemble modeling is the training efficiency. To evaluate the training efficiency of our framework, we compare EMKD with two best-performed baselines: MMInfoRec and DuoRec. Training efficiency is evaluated from two aspects: training speed (average training time required for one epoch), and convergence speed (time required to achieve satisfactory performance). To visualize the performance, we save the checkpoint from every epoch and test them afterwards. All experiments use the default hyper-parameters and are carried out on a single Tesla V100 GPU. 

Figure \ref{trainingefficiency_figure} presents the training efficiency of each model on Beauty and Toys datasets. We find out that EMKD with 2 parallel networks shows similar training speed (69.4s/epoch and 62.1s/epoch) with MMInfoRec (65.3s/epoch and 60.2s/epoch), while EMKD with 3 parallel networks (109.2s/epoch and 98.6s/epoch) is about 30\% slower than MMInfoRec. Besides, DuoRec shows the best training speed (40.8s/epoch and 37.9s/epoch) on both datasets.

We also find out that EMKD almost converges as fast as DuoRec. The performance curve of EMKD coincides with the performance curve of DuoRec around the first 1000 seconds, indicating that EMKD and DuoRec show similar convergence speed during this interval. However, the performance of EMKD continues to improve afterwards, while DuoRec reaches its best performance. By comparison, MMInfoRec takes about 4000 seconds to converge to a satisfactory performance, which is slower than DuoRec and EMKD.

Compared with DuoRec and MMInfoRec, EMKD only brings a minor reduction in training efficiency. However, the performance improvement can be very significant compared with training a single network. Therefore, we think that it is worthwhile to adopt ensemble modeling methods in sequential recommendation.

\subsection{Parameter Scaling (RQ6)}
EMKD trains multiple bidirectional Transformer encoders as an ensemble network, which will increase the total number of parameters. To determine whether the good performance of EMKD results from the proposed knowledge transfer mechanisms 
 or simply comes from the increase in the parameter size, we compare the performance of EMKD with other Transformer-based sequence encoders (SASRec and BERT4Rec) of similar parameter sizes. SASRec and BERT4Rec adopt two layers of Transformer blocks by default, and to scale the number of parameters, we stack more layers of Transformer blocks for SASRec and BERT4Rec so that they will have parameter sizes comparable to EMKD. Note that we count the parameters of both the embedding table and the sequence encoder(s) for each model, and we set the hidden dimensionality of each model as 256 for fair comparisons. Table \ref{parameterscalingtable} shows the parameter sizes and the performances of different models on three datasets. We can see that EMKD with 2 parallel networks has similar parameter size with a 6-layer SASRec or a 4-layer BERT4Rec, while EMKD with 3 parallel networks has similar parameter size with a 10-layer BERT4Rec. However, EMKD consistently outperforms SASRec and BERT4Rec of any parameter sizes, which verifies the effectiveness of the knowledge transfer mechanisms between the ensemble networks. We also find out that stacking more transformer blocks will not improve performance or even hurt performance in some cases, indicating that simply increasing the parameter size will not necessarily result in better performance for sequential recommendation.

\section{Conclusion}
In this paper, we present a novel framework called Ensemble Modeling with contrastive Knowledge Distillation for sequential recommendation (EMKD). Our framework adopts an ensemble of parallel networks to improve the overall prediction accuracy. To facilitate knowledge transfer between parallel networks, we propose a novel contrastive knowledge distillation approach that transfers knowledge from both the representation level and the logits level. We also design masked item prediction as the main task and attribute prediction as the auxiliary task for multi-task learning. Experiment results on three public benchmark datasets show that EMKD significantly outperforms baseline methods.
\begin{acks}
This research is partially supported by the NSFC (61876117, 62176175), the major project of natural science research in Universities of Jiangsu Province (21KJA520004), Suzhou Science and Technology Development Program (SYC2022139), the Priority Academic Program Development of Jiangsu Higher Education Institutions.
\end{acks}



\bibliographystyle{ACM-Reference-Format}
\balance
\bibliography{sample-base}


\begin{thebibliography}{45}


\ifx \showCODEN    \undefined \def \showCODEN     #1{\unskip}     \fi
\ifx \showDOI      \undefined \def \showDOI       #1{#1}\fi
\ifx \showISBNx    \undefined \def \showISBNx     #1{\unskip}     \fi
\ifx \showISBNxiii \undefined \def \showISBNxiii  #1{\unskip}     \fi
\ifx \showISSN     \undefined \def \showISSN      #1{\unskip}     \fi
\ifx \showLCCN     \undefined \def \showLCCN      #1{\unskip}     \fi
\ifx \shownote     \undefined \def \shownote      #1{#1}          \fi
\ifx \showarticletitle \undefined \def \showarticletitle #1{#1}   \fi
\ifx \showURL      \undefined \def \showURL       {\relax}        \fi
\providecommand\bibfield[2]{#2}
\providecommand\bibinfo[2]{#2}
\providecommand\natexlab[1]{#1}
\providecommand\showeprint[2][]{arXiv:#2}

\bibitem[Chen et~al\mbox{.}(2021a)]%
        {DistillingKnowledgeviaKnowledgeReview}
\bibfield{author}{\bibinfo{person}{Pengguang Chen}, \bibinfo{person}{Shu Liu},
  \bibinfo{person}{Hengshuang Zhao}, {and} \bibinfo{person}{Jiaya Jia}.}
  \bibinfo{year}{2021}\natexlab{a}.
\newblock \showarticletitle{Distilling Knowledge via Knowledge Review}. In
  \bibinfo{booktitle}{\emph{CVPR}}. \bibinfo{pages}{5008--5017}.
\newblock


\bibitem[Chen et~al\mbox{.}(2020b)]%
        {chen2020simple}
\bibfield{author}{\bibinfo{person}{Ting Chen}, \bibinfo{person}{Simon
  Kornblith}, \bibinfo{person}{Mohammad Norouzi}, {and}
  \bibinfo{person}{Geoffrey Hinton}.} \bibinfo{year}{2020}\natexlab{b}.
\newblock \showarticletitle{A Simple Framework for Contrastive Learning of
  Visual Representations}. In \bibinfo{booktitle}{\emph{ICML}}.
  \bibinfo{pages}{1597--1607}.
\newblock


\bibitem[Chen et~al\mbox{.}(2020a)]%
        {mocov2}
\bibfield{author}{\bibinfo{person}{Xinlei Chen}, \bibinfo{person}{Haoqi Fan},
  \bibinfo{person}{Ross Girshick}, {and} \bibinfo{person}{Kaiming He}.}
  \bibinfo{year}{2020}\natexlab{a}.
\newblock \showarticletitle{Improved Baselines with Momentum Contrastive
  Learning}. In \bibinfo{booktitle}{\emph{arXiv preprint arXiv:2003.04297}}.
\newblock


\bibitem[Chen and He(2021)]%
        {SimSiam}
\bibfield{author}{\bibinfo{person}{Xinlei Chen} {and} \bibinfo{person}{Kaiming
  He}.} \bibinfo{year}{2021}\natexlab{}.
\newblock \showarticletitle{Exploring Simple Siamese Representation Learning}.
  In \bibinfo{booktitle}{\emph{CVPR}}. \bibinfo{pages}{15745--15753}.
\newblock


\bibitem[Chen et~al\mbox{.}(2021b)]%
        {mocov3}
\bibfield{author}{\bibinfo{person}{Xinlei Chen}, \bibinfo{person}{Saining Xie},
  {and} \bibinfo{person}{Kaiming He}.} \bibinfo{year}{2021}\natexlab{b}.
\newblock \showarticletitle{An Empirical Study of Training Self-Supervised
  Vision Transformers}. In \bibinfo{booktitle}{\emph{ICCV}}.
  \bibinfo{pages}{9620--9629}.
\newblock


\bibitem[Devlin et~al\mbox{.}(2019)]%
        {Devlin2019BERT}
\bibfield{author}{\bibinfo{person}{Jacob Devlin}, \bibinfo{person}{Ming-Wei
  Chang}, \bibinfo{person}{Kenton Lee}, {and} \bibinfo{person}{Kristina
  Toutanova}.} \bibinfo{year}{2019}\natexlab{}.
\newblock \showarticletitle{{BERT}: Pre-training of Deep Bidirectional
  Transformers for Language Understanding}. In
  \bibinfo{booktitle}{\emph{NAACL}}. \bibinfo{pages}{4171--4186}.
\newblock


\bibitem[Fan et~al\mbox{.}(2022)]%
        {STOSA}
\bibfield{author}{\bibinfo{person}{Ziwei Fan}, \bibinfo{person}{Zhiwei Liu},
  \bibinfo{person}{Yu Wang}, \bibinfo{person}{Alice Wang},
  \bibinfo{person}{Zahra Nazari}, \bibinfo{person}{Lei Zheng},
  \bibinfo{person}{Hao Peng}, {and} \bibinfo{person}{Philip~S. Yu}.}
  \bibinfo{year}{2022}\natexlab{}.
\newblock \showarticletitle{Sequential Recommendation via Stochastic
  Self-Attention}. In \bibinfo{booktitle}{\emph{WWW}}.
  \bibinfo{pages}{2036--2047}.
\newblock


\bibitem[Fang et~al\mbox{.}(2021)]%
        {fang2021seed}
\bibfield{author}{\bibinfo{person}{Zhiyuan Fang}, \bibinfo{person}{Jianfeng
  Wang}, \bibinfo{person}{Lijuan Wang}, \bibinfo{person}{Lei Zhang},
  \bibinfo{person}{Yezhou Yang}, {and} \bibinfo{person}{Zicheng Liu}.}
  \bibinfo{year}{2021}\natexlab{}.
\newblock \showarticletitle{SEED: Self-supervised Distillation for Visual
  Representation}. In \bibinfo{booktitle}{\emph{ICLR}}.
\newblock


\bibitem[Gao et~al\mbox{.}(2021)]%
        {gao2021simcse}
\bibfield{author}{\bibinfo{person}{Tianyu Gao}, \bibinfo{person}{Xingcheng
  Yao}, {and} \bibinfo{person}{Danqi Chen}.} \bibinfo{year}{2021}\natexlab{}.
\newblock \showarticletitle{SimCSE: Simple Contrastive Learning of Sentence
  Embeddings}. In \bibinfo{booktitle}{\emph{EMNLP}}.
  \bibinfo{pages}{6894--6910}.
\newblock


\bibitem[Garipov et~al\mbox{.}(2018)]%
        {fastdnnensemble}
\bibfield{author}{\bibinfo{person}{Timur Garipov}, \bibinfo{person}{Pavel
  Izmailov}, \bibinfo{person}{Dmitrii Podoprikhin}, \bibinfo{person}{Dmitry~P
  Vetrov}, {and} \bibinfo{person}{Andrew~G Wilson}.}
  \bibinfo{year}{2018}\natexlab{}.
\newblock \showarticletitle{Loss Surfaces, Mode Connectivity, and Fast
  Ensembling of DNNs}. In \bibinfo{booktitle}{\emph{NeurIPS}}.
  \bibinfo{pages}{8803–8812}.
\newblock


\bibitem[Han et~al\mbox{.}(2018)]%
        {CoTeaching}
\bibfield{author}{\bibinfo{person}{Bo Han}, \bibinfo{person}{Quanming Yao},
  \bibinfo{person}{Xingrui Yu}, \bibinfo{person}{Gang Niu},
  \bibinfo{person}{Miao Xu}, \bibinfo{person}{Weihua Hu},
  \bibinfo{person}{Ivor~W. Tsang}, {and} \bibinfo{person}{Masashi Sugiyama}.}
  \bibinfo{year}{2018}\natexlab{}.
\newblock \showarticletitle{Co-teaching: Robust training of deep neural
  networks with extremely noisy labels}. In
  \bibinfo{booktitle}{\emph{NeurIPS}}. \bibinfo{pages}{8536--8546}.
\newblock


\bibitem[Harper and Konstan(2015)]%
        {movielens}
\bibfield{author}{\bibinfo{person}{F.~Maxwell Harper} {and}
  \bibinfo{person}{Joseph~A. Konstan}.} \bibinfo{year}{2015}\natexlab{}.
\newblock \showarticletitle{The MovieLens Datasets: History and Context}.
\newblock \bibinfo{journal}{\emph{ACM Trans. Interact. Intell. Syst.}}
  (\bibinfo{year}{2015}), \bibinfo{pages}{1--19}.
\newblock


\bibitem[He et~al\mbox{.}(2020)]%
        {mocov1}
\bibfield{author}{\bibinfo{person}{Kaiming He}, \bibinfo{person}{Haoqi Fan},
  \bibinfo{person}{Yuxin Wu}, \bibinfo{person}{Saining Xie}, {and}
  \bibinfo{person}{Ross Girshick}.} \bibinfo{year}{2020}\natexlab{}.
\newblock \showarticletitle{Momentum Contrast for Unsupervised Visual
  Representation Learning}. In \bibinfo{booktitle}{\emph{CVPR}}.
  \bibinfo{pages}{9726--9735}.
\newblock


\bibitem[He and McAuley(2016)]%
        {Fossil}
\bibfield{author}{\bibinfo{person}{Ruining He} {and} \bibinfo{person}{Julian
  McAuley}.} \bibinfo{year}{2016}\natexlab{}.
\newblock \showarticletitle{Fusing Similarity Models with Markov Chains for
  Sparse Sequential Recommendation}. In \bibinfo{booktitle}{\emph{ICDM}}.
  \bibinfo{pages}{191--200}.
\newblock


\bibitem[Hidasi et~al\mbox{.}(2016)]%
        {srnn2016}
\bibfield{author}{\bibinfo{person}{Bal{\'{a}}zs Hidasi},
  \bibinfo{person}{Alexandros Karatzoglou}, \bibinfo{person}{Linas Baltrunas},
  {and} \bibinfo{person}{Domonkos Tikk}.} \bibinfo{year}{2016}\natexlab{}.
\newblock \showarticletitle{Session-based Recommendations with Recurrent Neural
  Networks}.
\newblock \bibinfo{journal}{\emph{ICLR}}.
\newblock


\bibitem[Hinton et~al\mbox{.}(2015)]%
        {KnowledgeDistillation}
\bibfield{author}{\bibinfo{person}{Geoffrey Hinton}, \bibinfo{person}{Oriol
  Vinyals}, {and} \bibinfo{person}{Jeff Dean}.}
  \bibinfo{year}{2015}\natexlab{}.
\newblock \showarticletitle{Distilling the Knowledge in a Neural Network}.
\newblock \bibinfo{journal}{\emph{arXiv preprint arXiv:1503.02531}}
  (\bibinfo{year}{2015}).
\newblock


\bibitem[Huang et~al\mbox{.}(2017)]%
        {snapshotensemble}
\bibfield{author}{\bibinfo{person}{Gao Huang}, \bibinfo{person}{Yixuan Li},
  \bibinfo{person}{Geoff Pleiss}, \bibinfo{person}{Zhuang Liu},
  \bibinfo{person}{John~E. Hopcroft}, {and} \bibinfo{person}{Kilian~Q.
  Weinberger}.} \bibinfo{year}{2017}\natexlab{}.
\newblock \showarticletitle{Snapshot Ensembles: Train 1, Get {M} for Free}. In
  \bibinfo{booktitle}{\emph{ICLR}}.
\newblock


\bibitem[Kang and McAuley(2018)]%
        {kang18attentive}
\bibfield{author}{\bibinfo{person}{Wang-Cheng Kang} {and}
  \bibinfo{person}{Julian McAuley}.} \bibinfo{year}{2018}\natexlab{}.
\newblock \showarticletitle{Self-Attentive Sequential Recommendation}. In
  \bibinfo{booktitle}{\emph{ICDM}}. \bibinfo{pages}{197--206}.
\newblock


\bibitem[Kawaguchi(2016)]%
        {localminima}
\bibfield{author}{\bibinfo{person}{Kenji Kawaguchi}.}
  \bibinfo{year}{2016}\natexlab{}.
\newblock \showarticletitle{Deep Learning without Poor Local Minima}. In
  \bibinfo{booktitle}{\emph{NeurIPS}}. \bibinfo{pages}{586–594}.
\newblock


\bibitem[Kingma and Ba(2015)]%
        {Adam}
\bibfield{author}{\bibinfo{person}{Diederik~P. Kingma} {and}
  \bibinfo{person}{Jimmy Ba}.} \bibinfo{year}{2015}\natexlab{}.
\newblock \showarticletitle{Adam: {A} Method for Stochastic Optimization}. In
  \bibinfo{booktitle}{\emph{ICLR}}.
\newblock


\bibitem[Krichene and Rendle(2020)]%
        {krichene2020sampled}
\bibfield{author}{\bibinfo{person}{Walid Krichene} {and}
  \bibinfo{person}{Steffen Rendle}.} \bibinfo{year}{2020}\natexlab{}.
\newblock \showarticletitle{On Sampled Metrics for Item Recommendation}. In
  \bibinfo{booktitle}{\emph{SIGKDD}}. \bibinfo{pages}{1748--1757}.
\newblock


\bibitem[Lan et~al\mbox{.}(2018)]%
        {ensembleonthefly}
\bibfield{author}{\bibinfo{person}{Xu Lan}, \bibinfo{person}{Xiatian Zhu},
  {and} \bibinfo{person}{Shaogang Gong}.} \bibinfo{year}{2018}\natexlab{}.
\newblock \showarticletitle{Knowledge Distillation by On-the-Fly Native
  Ensemble}. In \bibinfo{booktitle}{\emph{NeurIPS}}.
  \bibinfo{pages}{7528–7538}.
\newblock


\bibitem[McAuley et~al\mbox{.}(2015)]%
        {mcauley15image}
\bibfield{author}{\bibinfo{person}{Julian McAuley},
  \bibinfo{person}{Christopher Targett}, \bibinfo{person}{Javen Shi}, {and}
  \bibinfo{person}{Anton van~den Hengel}.} \bibinfo{year}{2015}\natexlab{}.
\newblock \showarticletitle{Image-based Recommendations on Styles and
  Substitutes}. In \bibinfo{booktitle}{\emph{SIGIR}}. \bibinfo{pages}{43--52}.
\newblock


\bibitem[Paszke et~al\mbox{.}(2019)]%
        {pytorch}
\bibfield{author}{\bibinfo{person}{Adam Paszke}, \bibinfo{person}{Sam Gross},
  \bibinfo{person}{Francisco Massa}, \bibinfo{person}{Adam Lerer},
  \bibinfo{person}{James Bradbury}, \bibinfo{person}{Gregory Chanan},
  \bibinfo{person}{Trevor Killeen}, \bibinfo{person}{Zeming Lin},
  \bibinfo{person}{Natalia Gimelshein}, \bibinfo{person}{Luca Antiga},
  \bibinfo{person}{Alban Desmaison}, \bibinfo{person}{Andreas K\"{o}pf},
  \bibinfo{person}{Edward Yang}, \bibinfo{person}{Zach DeVito},
  \bibinfo{person}{Martin Raison}, \bibinfo{person}{Alykhan Tejani},
  \bibinfo{person}{Sasank Chilamkurthy}, \bibinfo{person}{Benoit Steiner},
  \bibinfo{person}{Lu Fang}, \bibinfo{person}{Junjie Bai}, {and}
  \bibinfo{person}{Soumith Chintala}.} \bibinfo{year}{2019}\natexlab{}.
\newblock \showarticletitle{PyTorch: An Imperative Style, High-Performance Deep
  Learning Library}. In \bibinfo{booktitle}{\emph{NeurIPS}}.
  \bibinfo{pages}{8024--8035}.
\newblock


\bibitem[Qiu et~al\mbox{.}(2021)]%
        {MMInfoRec}
\bibfield{author}{\bibinfo{person}{Ruihong Qiu}, \bibinfo{person}{Zi Huang},
  {and} \bibinfo{person}{Hongzhi Yin}.} \bibinfo{year}{2021}\natexlab{}.
\newblock \showarticletitle{Memory Augmented Multi-Instance Contrastive
  Predictive Coding for Sequential Recommendation}. In
  \bibinfo{booktitle}{\emph{ICDM}}. \bibinfo{pages}{519--528}.
\newblock


\bibitem[Qiu et~al\mbox{.}(2022)]%
        {DuoRec}
\bibfield{author}{\bibinfo{person}{Ruihong Qiu}, \bibinfo{person}{Zi Huang},
  \bibinfo{person}{Hongzhi Yin}, {and} \bibinfo{person}{Zijian Wang}.}
  \bibinfo{year}{2022}\natexlab{}.
\newblock \showarticletitle{Contrastive Learning for Representation
  Degeneration Problem in Sequential Recommendation}. In
  \bibinfo{booktitle}{\emph{WSDM}}. \bibinfo{pages}{813–823}.
\newblock


\bibitem[Rendle et~al\mbox{.}(2010)]%
        {FPMC}
\bibfield{author}{\bibinfo{person}{Steffen Rendle}, \bibinfo{person}{Christoph
  Freudenthaler}, {and} \bibinfo{person}{Lars Schmidt{-}Thieme}.}
  \bibinfo{year}{2010}\natexlab{}.
\newblock \showarticletitle{Factorizing personalized Markov chains for
  next-basket recommendation}. In \bibinfo{booktitle}{\emph{WWW}}.
  \bibinfo{pages}{811–820}.
\newblock


\bibitem[Shani et~al\mbox{.}(2005)]%
        {MDP}
\bibfield{author}{\bibinfo{person}{Guy Shani}, \bibinfo{person}{Ronen~I.
  Brafman}, {and} \bibinfo{person}{David Heckerman}.}
  \bibinfo{year}{2005}\natexlab{}.
\newblock \showarticletitle{An MDP-Based Recommender System}.
\newblock \bibinfo{journal}{\emph{JMLR}}, \bibinfo{pages}{1265–1295}.
\newblock


\bibitem[Song and Chai(2018)]%
        {CollaborativeLearning}
\bibfield{author}{\bibinfo{person}{Guocong Song} {and} \bibinfo{person}{Wei
  Chai}.} \bibinfo{year}{2018}\natexlab{}.
\newblock \showarticletitle{Collaborative Learning for Deep Neural Networks}.
  In \bibinfo{booktitle}{\emph{NeurIPS}}. \bibinfo{pages}{1837–1846}.
\newblock


\bibitem[Srivastava et~al\mbox{.}(2014)]%
        {dropout}
\bibfield{author}{\bibinfo{person}{Nitish Srivastava},
  \bibinfo{person}{Geoffrey Hinton}, \bibinfo{person}{Alex Krizhevsky},
  \bibinfo{person}{Ilya Sutskever}, {and} \bibinfo{person}{Ruslan
  Salakhutdinov}.} \bibinfo{year}{2014}\natexlab{}.
\newblock \showarticletitle{Dropout: A Simple Way to Prevent Neural Networks
  from Overfitting}.
\newblock \bibinfo{journal}{\emph{JMLR}} (\bibinfo{year}{2014}),
  \bibinfo{pages}{1929--1958}.
\newblock


\bibitem[Sun et~al\mbox{.}(2019)]%
        {Sun2019bert}
\bibfield{author}{\bibinfo{person}{Fei Sun}, \bibinfo{person}{Jun Liu},
  \bibinfo{person}{Jian Wu}, \bibinfo{person}{Changhua Pei},
  \bibinfo{person}{Xiao Lin}, \bibinfo{person}{Wenwu Ou}, {and}
  \bibinfo{person}{Peng Jiang}.} \bibinfo{year}{2019}\natexlab{}.
\newblock \showarticletitle{BERT4Rec: Sequential Recommendation with
  Bidirectional Encoder Representations from Transformer}. In
  \bibinfo{booktitle}{\emph{CIKM}}. \bibinfo{pages}{1441–1450}.
\newblock


\bibitem[Tang and Wang(2018)]%
        {tang2018caser}
\bibfield{author}{\bibinfo{person}{Jiaxi Tang} {and} \bibinfo{person}{Ke
  Wang}.} \bibinfo{year}{2018}\natexlab{}.
\newblock \showarticletitle{Personalized Top-N Sequential Recommendation via
  Convolutional Sequence Embedding}. In \bibinfo{booktitle}{\emph{WSDM}}.
  \bibinfo{pages}{565–573}.
\newblock


\bibitem[Tian et~al\mbox{.}(2020)]%
        {ContrastiveRepresentationDistillation}
\bibfield{author}{\bibinfo{person}{Yonglong Tian}, \bibinfo{person}{Dilip
  Krishnan}, {and} \bibinfo{person}{Phillip Isola}.}
  \bibinfo{year}{2020}\natexlab{}.
\newblock \showarticletitle{Contrastive Representation Distillation}. In
  \bibinfo{booktitle}{\emph{ICLR}}.
\newblock


\bibitem[Tramèr et~al\mbox{.}(2018)]%
        {ensembleadversarialtraining}
\bibfield{author}{\bibinfo{person}{Florian Tramèr}, \bibinfo{person}{Alexey
  Kurakin}, \bibinfo{person}{Nicolas Papernot}, \bibinfo{person}{Ian
  Goodfellow}, \bibinfo{person}{Dan Boneh}, {and} \bibinfo{person}{Patrick
  McDaniel}.} \bibinfo{year}{2018}\natexlab{}.
\newblock \showarticletitle{Ensemble Adversarial Training: Attacks and
  Defenses}. In \bibinfo{booktitle}{\emph{ICLR}}.
\newblock


\bibitem[van~den Oord et~al\mbox{.}(2018)]%
        {contrastivepredictivecoding}
\bibfield{author}{\bibinfo{person}{A{\"{a}}ron van~den Oord},
  \bibinfo{person}{Yazhe Li}, {and} \bibinfo{person}{Oriol Vinyals}.}
  \bibinfo{year}{2018}\natexlab{}.
\newblock \showarticletitle{Representation Learning with Contrastive Predictive
  Coding}.
\newblock \bibinfo{journal}{\emph{arXiv preprint arXiv:1807.03748}}
  (\bibinfo{year}{2018}).
\newblock


\bibitem[Vaswani et~al\mbox{.}(2017)]%
        {vaswani2017transformer}
\bibfield{author}{\bibinfo{person}{Ashish Vaswani}, \bibinfo{person}{Noam
  Shazeer}, \bibinfo{person}{Niki Parmar}, \bibinfo{person}{Jakob Uszkoreit},
  \bibinfo{person}{Llion Jones}, \bibinfo{person}{Aidan~N Gomez},
  \bibinfo{person}{{\L}ukasz Kaiser}, {and} \bibinfo{person}{Illia
  Polosukhin}.} \bibinfo{year}{2017}\natexlab{}.
\newblock \showarticletitle{Attention is All you Need}. In
  \bibinfo{booktitle}{\emph{NeurIPS}}. \bibinfo{pages}{6000–6010}.
\newblock


\bibitem[Wang and Liu(2021)]%
        {wang2021understanding}
\bibfield{author}{\bibinfo{person}{Feng Wang} {and} \bibinfo{person}{Huaping
  Liu}.} \bibinfo{year}{2021}\natexlab{}.
\newblock \showarticletitle{Understanding the Behaviour of Contrastive Loss}.
  In \bibinfo{booktitle}{\emph{CVPR}}. \bibinfo{pages}{2495--2504}.
\newblock


\bibitem[Wang and Isola(2020)]%
        {wang2020alignmentuniformity}
\bibfield{author}{\bibinfo{person}{Tongzhou Wang} {and}
  \bibinfo{person}{Phillip Isola}.} \bibinfo{year}{2020}\natexlab{}.
\newblock \showarticletitle{Understanding Contrastive Representation Learning
  Through Alignment and Uniformity on the Hypersphere}. In
  \bibinfo{booktitle}{\emph{ICML}}. \bibinfo{pages}{9929--9939}.
\newblock


\bibitem[Wei et~al\mbox{.}(2022)]%
        {CML}
\bibfield{author}{\bibinfo{person}{Wei Wei}, \bibinfo{person}{Chao Huang},
  \bibinfo{person}{Lianghao Xia}, \bibinfo{person}{Yong Xu},
  \bibinfo{person}{Jiashu Zhao}, {and} \bibinfo{person}{Dawei Yin}.}
  \bibinfo{year}{2022}\natexlab{}.
\newblock \showarticletitle{Contrastive Meta Learning with Behavior
  Multiplicity for Recommendation}. In \bibinfo{booktitle}{\emph{WSDM}}.
  \bibinfo{pages}{1120–1128}.
\newblock


\bibitem[Xie et~al\mbox{.}(2022)]%
        {Xu2020Contrastive}
\bibfield{author}{\bibinfo{person}{Xu Xie}, \bibinfo{person}{Fei Sun},
  \bibinfo{person}{Zhaoyang Liu}, \bibinfo{person}{Shiwen Wu},
  \bibinfo{person}{Jinyang Gao}, \bibinfo{person}{Jiandong Zhang},
  \bibinfo{person}{Bolin Ding}, {and} \bibinfo{person}{Bin Cui}.}
  \bibinfo{year}{2022}\natexlab{}.
\newblock \showarticletitle{Contrastive Learning for Sequential
  Recommendation}. In \bibinfo{booktitle}{\emph{ICDE}}.
  \bibinfo{pages}{1259--1273}.
\newblock


\bibitem[Yan et~al\mbox{.}(2021)]%
        {yan-etal-2021-consert}
\bibfield{author}{\bibinfo{person}{Yuanmeng Yan}, \bibinfo{person}{Rumei Li},
  \bibinfo{person}{Sirui Wang}, \bibinfo{person}{Fuzheng Zhang},
  \bibinfo{person}{Wei Wu}, {and} \bibinfo{person}{Weiran Xu}.}
  \bibinfo{year}{2021}\natexlab{}.
\newblock \showarticletitle{{C}on{SERT}: A Contrastive Framework for
  Self-Supervised Sentence Representation Transfer}. In
  \bibinfo{booktitle}{\emph{ACL}}. \bibinfo{pages}{5065--5075}.
\newblock


\bibitem[Yu et~al\mbox{.}(2022)]%
        {simgcl}
\bibfield{author}{\bibinfo{person}{Junliang Yu}, \bibinfo{person}{Hongzhi Yin},
  \bibinfo{person}{Xin Xia}, \bibinfo{person}{Tong Chen},
  \bibinfo{person}{Lizhen Cui}, {and} \bibinfo{person}{Quoc Viet~Hung Nguyen}.}
  \bibinfo{year}{2022}\natexlab{}.
\newblock \showarticletitle{Are Graph Augmentations Necessary? Simple Graph
  Contrastive Learning for Recommendation}. In
  \bibinfo{booktitle}{\emph{SIGIR}}. \bibinfo{pages}{1294–1303}.
\newblock


\bibitem[Zhang et~al\mbox{.}(2019)]%
        {FDSA}
\bibfield{author}{\bibinfo{person}{Tingting Zhang}, \bibinfo{person}{Pengpeng
  Zhao}, \bibinfo{person}{Yanchi Liu}, \bibinfo{person}{Victor~S. Sheng},
  \bibinfo{person}{Jiajie Xu}, \bibinfo{person}{Deqing Wang},
  \bibinfo{person}{Guanfeng Liu}, {and} \bibinfo{person}{Xiaofang Zhou}.}
  \bibinfo{year}{2019}\natexlab{}.
\newblock \showarticletitle{Feature-level Deeper Self-Attention Network for
  Sequential Recommendation}. In \bibinfo{booktitle}{\emph{IJCAI}}.
  \bibinfo{pages}{4320--4326}.
\newblock


\bibitem[Zhang et~al\mbox{.}(2018)]%
        {DeepMutualLearning}
\bibfield{author}{\bibinfo{person}{Ying Zhang}, \bibinfo{person}{Tao Xiang},
  \bibinfo{person}{Timothy~M. Hospedales}, {and} \bibinfo{person}{Huchuan Lu}.}
  \bibinfo{year}{2018}\natexlab{}.
\newblock \showarticletitle{Deep Mutual Learning}. In
  \bibinfo{booktitle}{\emph{CVPR}}. \bibinfo{pages}{4320--4328}.
\newblock


\bibitem[Zhou et~al\mbox{.}(2020)]%
        {CIKM2020-S3Rec}
\bibfield{author}{\bibinfo{person}{Kun Zhou}, \bibinfo{person}{Hui Wang},
  \bibinfo{person}{Wayne~Xin Zhao}, \bibinfo{person}{Yutao Zhu},
  \bibinfo{person}{Sirui Wang}, \bibinfo{person}{Fuzheng Zhang},
  \bibinfo{person}{Zhongyuan Wang}, {and} \bibinfo{person}{Ji{-}Rong Wen}.}
  \bibinfo{year}{2020}\natexlab{}.
\newblock \showarticletitle{S3-Rec: Self-Supervised Learning for Sequential
  Recommendation with Mutual Information Maximization}. In
  \bibinfo{booktitle}{\emph{CIKM}}. \bibinfo{pages}{1893–1902}.
\newblock


\end{thebibliography}
\appendix


\end{document}